\documentclass{article}
\usepackage{color}
\usepackage{amsfonts}
\usepackage{rlepsf}
\usepackage{graphics}
\usepackage{amssymb}  
\usepackage{amsmath}
\usepackage{amssymb}
\usepackage{amscd}
\usepackage{euscript}
\usepackage{enumerate}
\usepackage[all]{xy}
\usepackage{fullpage}

\newtheorem{Theorem}{Theorem}
\newtheorem{Corollary}[Theorem]{Corollary}
\newtheorem{Example}[Theorem]{Example}
\newtheorem{Definition}[Theorem]{Definition}
\newtheorem{Remark}[Theorem]{Remark}

\newtheorem{Lemma}[Theorem]{Lemma}
\newtheorem{Proposition}[Theorem]{Proposition}

\newtheorem{Problem}{Problem}

\newenvironment{Proof}[1][Proof]{\textbf{#1.} }{ \rule{0.5em}{0.5em}}

\newcommand{\h}{\mathfrak{h}}
\newcommand{\g}{\mathfrak{g}}

\def \coker {\rm coker}

\def \s {\sigma}
\def \btn {\,\overline{\otimes}\,}

\def \lX {\mathfrak{G}}

\def \Wc {\mathcal{W}}
\def \k {\mathfrak{k}}

\def \S {\EuScript{S}}
\def \GL {{\rm GL}}

\def \U {\EuScript{U}}
\def \s {\sigma}
\def \V {\EuScript{V}}

\def \Wc  {\mathcal{W}}
\def \C {\mathbb{C}}

\def \Der {\mathrm{Der}}

\def \e {\mathfrak{h}}
\def \w {\omega}
\def \id {\mathrm{id}}
\def \g {\mathfrak {g}}

\def \f {\phi}
\def \d {\partial}
\def \M {\EuScript{M}}

\def \b {\beta}
\def \g {\mathfrak{g}}

\def \tn {\otimes}
\def \t {\triangleright}

\def \ra {\xrightarrow}
\def \le {\mathfrak{h}}

\def \Wc {{\EuScript{W}}}

\def \Z {\mathbb{Z}}

\def \fo {\textrm{ for  each }}
\def \ane {\textrm{ and each } }
\def \an {\textrm{ and }}
\def \GL {\mathrm{GL}}
\def \gl {\mathfrak{gl}}
\def \F {\mathrm{F}}
\def \G {\EuScript{G}}

\def \an {\textrm{ and }}
\def \Hom {\mathrm{Hom}}
\def\be{\begin{equation}}
\def\ee{\end{equation}}
\def\bea{\begin{eqnarray}}
\def\eea{\end{eqnarray}}

\def \Aut {\mathrm{Aut}}
\def \Ad {\mathrm{Ad}}
\def \h {\mathfrak{h}}
\def \act {\triangleright}
\def \Pf {\mathsf{Pf}}
\def \pf {\mathfrak{pf}}
\def \fxmod {\mathsf{FdX}(S,\d_0)}
\def \ch {\mathsf{ch}}
\def \ftch {\mathsf{f}2\mathsf{ch}}
\def \tch {2\mathsf{ch}}
\def \tchmf {2\mathfrak{ch}}
\def \Uk {\bar{\mathfrak{U}}^{(k)}}

\def \Tk {\mathfrak{T}^{(k)}}

\begin{document}

\title{Categorifying the Knizhnik-Zamolodchikov Connection}

\author{ Lucio Simone Cirio \footnote{Current address: Mathematics Research Unit, University of Luxembourg, 6, rue Richard Coudenhove-Kalergi, L-1359 Luxembourg, Luxembourg. Email: {\it \small lucio.cirio@uni.lu}} \\ 
{\footnotesize Grupo de Fisica Matem\'{a}tica da Universidade de Lisboa (GFM-UL) } \\ 
{\footnotesize Instituto para a Investiga\c{c}\~{a}o Interdisciplinar } \\ 
{\footnotesize Av. Prof. Gama Pinto 2, 1649-003 Lisboa, Portugal} \\
{\it \small cirio@cii.fc.ul.pt} \vspace{.4cm} \\ 
Jo\~{a}o  Faria Martins  \\
{\footnotesize Departamento de Matem\'{a}tica } \\ 
{\footnotesize Faculdade de Ci\^{e}ncias e Tecnologia,   Universidade Nova de Lisboa} \\
{\footnotesize  Quinta da Torre, 2829-516 Caparica, Portugal } \\
{\it \small jn.martins@fct.unl.pt } }

\date{}
\maketitle

\begin{abstract}
\noindent
{In the context of higher gauge theory, we construct a flat and fake flat 2-connection,  in the configuration space of $n$ particles in the complex plane, categorifying the Knizhnik-Zamolodchikov connection. To this end, we define  the differential crossed module of  horizontal 2-chord diagrams, categorifying the Lie algebra of horizontal chord diagrams in a set of $n$ parallel copies of the interval. This therefore yields a categorification of the 4-term relation. We carefully discuss the representation theory of differential crossed modules in chain-complexes of vector spaces, which makes it possible to formulate the notion of an infinitesimal 2-R matrix in a differential crossed module.}
\end{abstract}

\vspace{.5cm}
\noindent
{\it keyword:} higher gauge theory, braided surface, two-dimensional holonomy, chord diagrams, infinitesimal braiding, 4-term relation, differential crossed module, Knizhnik-Zamolodchikov equations, categorical representation. \\[.2cm]
{\it MSC2010:} {{16T25, 
20F36 
(principal);  
18D05, 
17B37, 
53C29, 
57Q45 
(secondary).}}


\section{Introduction, Motivation and Background}

\begin{figure}[h!] 
{\centerline{\relabelbox 
\epsfysize 2.5cm
\epsfbox{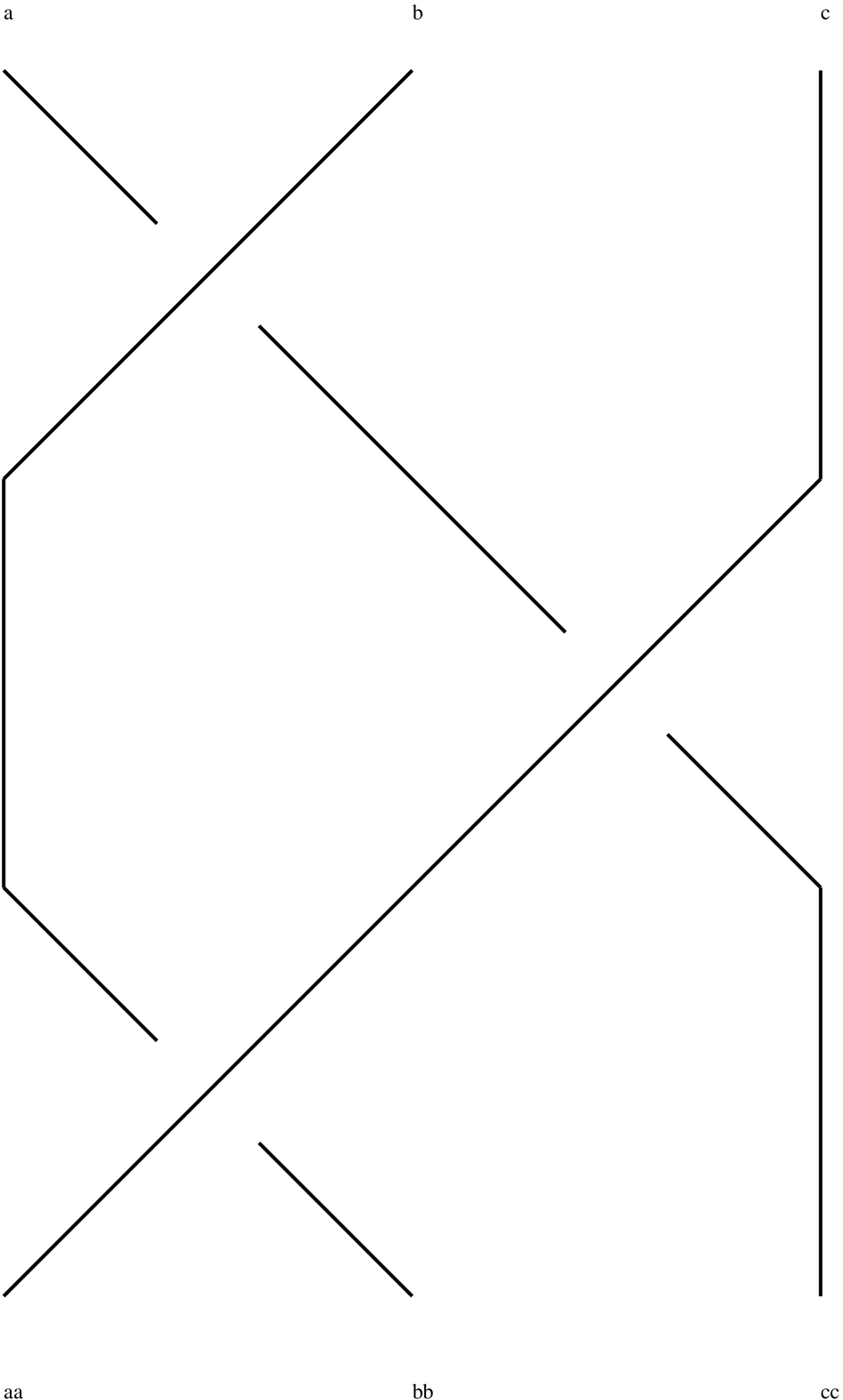}
\relabel{a}{$\scriptscriptstyle{1}$}
\relabel{b}{$\scriptscriptstyle{2}$}
\relabel{c}{$\scriptscriptstyle{3}$}
\relabel{aa}{$\scriptscriptstyle{1}$}
\relabel{bb}{$\scriptscriptstyle{2}$}
\relabel{cc}{$\scriptscriptstyle{3}$}
\endrelabelbox}}
\caption{{\label{braid} A braid with three strands.}}
\end{figure}

Let $I=[0,1]$. Given a positive integer $n$, a braid \cite{Bi,BiBr,Ka}  with $n$-strands {$b=\{x_i(t)\}_{i=1}^n$}  is, by definition, a (piecewise smooth, neat) embedding of the manifold $I\sqcup \dots \sqcup I=I^{\sqcup n}$ into $\C\times I$, such that for any $i$ the projection of $x_i(t)$ in the last variable is monotone. In addition we suppose that for every $i$ we have {$x_i(0)\in \{1,\dots,n\}\times \{0\}$ and $x_i(1)\in \{1,\dots,n\}\times \{1\}$}.
Braids are considered equivalent if they differ by a boundary preserving ambient isotopy. Two braids $b$ and $b'$ with $n$-strands can be multiplied by placing $b$ on top of $b'$ in the obvious way. This defines a group $B_n$ called the Artin braid group \cite{Art} with $n$-strands. This is the group with generators $X_{i}$, where $i\in \{1,\dots,n-1\}$ and relations
 \begin{align}
X_{i}X_{i+i}X_i&=X_{i+1}X_iX_{i+1}, \textrm{ if } i\in \{1,\dots, n-2\}\label{br1}\\
X_i X_j&=X_jX_i, \textrm{ if } |i-j| \ge 2 \textrm{ and } i,j\in \{1,\dots, n-1\} \label{br2}.
 \end{align}
The braid in figure \ref{braid} is given by $X_1X_2X_1$ in terms of these generators.

\noindent
There is an obvious group morphism $p\colon B_n \to S_n$ from $B_n$ onto the symmetric group $S_n$ of symmetries of the set $\{1,\dots,n\}$. The pure braid group $P_n$ is by definition the kernel of this map. 

Let $n$ be a positive integer. The space $\C(n)$ of $n$ distinguishable particles in the complex plane $\C$ is by definition the manifold on $n$-tuples $(z_1,\dots, z_n) \in\C^n$ such that $z_i \neq z_j$ if $i \neq j$. This is an aspherical manifold. There exists an obvious action of $S_n$ on $\C(n)$ by permuting coordinates. The space of $n$ {indistinguishable} particles in the complex plane is defined as $\C(n)/S_n$. Since $\C(n)$ is aspherical then so is $\C(n)/S_n$.

It is well known, and not difficult to  see, that the pure braid group $P_n$ is isomorphic to the fundamental group of $\C(n)$, and that the braid group $B_n$ is isomorphic to the fundamental group of $\C(n)/S_n$. Proofs are in \cite{Bi,Ka}.

Let us be given  a Lie algebra $\g$ with a $\C$-valued $\g$-invariant non-degenerate symmetric bilinear form ${\langle-,- \rangle}$. Let $r=\sum_{i} t_i \tn s_i\in \g \tn \g$ be the associated tensor; that is $\langle X,Y \rangle=\sum_i  \langle X,s_i\rangle\langle Y, t_i\rangle,$ for each $X,Y \in \g$. Choose a representation of $\g$ on a vector space $V$, with action denoted by $x \t v$, with $x \in \g$ and $v \in V$. 
Denote the tensor product $V\tn \dots \tn V$ of $V$ with itself $n$ times as $V^{\tn n}$, and the Lie algebra of linear maps $V^{\tn n} \to V^{\tn n}$ by $\Hom(V^{\tn n})$. Consider the trivial vector bundle $\C(n) \times V^{\tn n}$. The Knizhnik-Zamolodchikov connection (KZ-connection)  is given by the following $\Hom(V^{\tn n})$-valued form in the configuration space $\C(n)$
$$A=\frac{h}{2 \pi i}\sum_{a <b} \w_{ab} \phi_{ab}(r),$$ 
where $a,b \in \{1,\dots,n\}$,  $\w_{ab}=\frac{d z_a -d z_b}{z_a-z_b}$ and $\phi_{ab}(r)\colon   V^{\tn n} \to  V^{\tn n}$ is the linear map (we call it insertion map) such that $$\phi_{ab}(r)(v_1 \tn  \dots \tn v_a \tn \dots \tn v_b\tn \dots\tn v_n)=\sum_{i} v_1 \tn  \dots \tn s_i\t v_a \tn \dots \tn t_i \t v_b\tn \dots\tn v_n.$$  
This connection  appeared originally in the context of conformal field theory \cite{KZ}, being also natural in the context of the quantization of the Chern-Simons action \cite{Wi}; see also the books \cite{HW,Kh2}.

We have actions of $S_n$ on $\C(n)$ and of $S_n$ on  $V^{\tn n}$, and therefore the product action of $S_n$ in $\C(n) \times V^{\tn n}$ is an action by vector bundle maps. Consider the quotient vector bundle $(\C(n) \times V^{\tn n})/S_n$, over  $\C(n)/S_n$.
 Since, clearly, the KZ-connection is invariant under this action, we also have a quotient connection $A$ on the vector bundle $(\C(n) \times V^{\tn n})/S_n$. This connection will also be called the KZ-connection.

The KZ-connection $A$, both in $\C(n)$ and in $\C(n)/S_n$  is flat, in other words the curvature 2-form $dA+\frac{1}{2} A\wedge A$ vanishes.  This follows from the $\g$-invariance of the (symmetric and non-degenerate) bilinear form ${\langle-,- \rangle}$, which implies the relation 
\begin{equation}\label{4t}
[r_{12}+r_{13},r_{23}]=0, \textrm{ in } \g \tn \g \tn \g \subset \U(\g)\tn \U(\g) \tn \U(\g),
\end{equation}
where $\U(\g)$ is the universal enveloping algebra of $\g$. For $r=\sum_i s_i \tn t_i$, we have put
\begin{align} \label{rab}
 r_{12}&=\sum_{i} s_i \tn t_i \tn 1
 &r_{13}&=\sum_{i} s_i  \tn 1\tn t_i 
& r_{23}&=\sum_{i} 1\tn s_i  \tn t_i.
\end{align}
Relation \eqref{4t} is  called the 4-term relation, \cite{BN}. At the level of insertion maps it implies that: 
\begin{align}\label{ibr1} 
 \phi_{ab}(r)\phi_{bc}(r)+\phi_{ac}(r)\phi_{bc}(r)=\phi_{bc}(r)\phi_{ab}(r)+\phi_{bc}(r)\phi_{ac}(r),
\end{align}  for each $a,b,c \in \{1,\dots ,n\}$.
We also have rather obviously:
\begin{align}\label{ibr2} 
[\phi_{ab}(r),\phi_{a'b'}(r)]=0, \textrm{ if } \{a,b\} \cap \{a',b'\}=\emptyset. 
\end{align}
Relations \eqref{ibr1} and \eqref{ibr2} are called infinitesimal braid relations, being an infinitesimal counterpart of the braid group relations \eqref{br1} and \eqref{br2}.

We therefore define an {\it infinitesimal R-matrix} in an arbitrary Lie algebra $\g$ as being an arbitrary symmetric tensor $r \in \g \tn \g$ satisfying the 4-term relation \eqref{4t}. Any infinitesimal R-matrix in $\g$ yields a flat connection $A=\frac{h}{2 \pi i}\sum_{a <b} \w_{ab} \phi_{ab}(r)$ in $\C(n)$, for any representation $V$ of $\g$.

The flatness of $A$, and the fact that $A$ is invariant under the action of the symmetric group $S_n$, implies in particular that the holonomy of $A$ descends to a group morphism $B_n =\pi_1(\C(n)/S_n)\to \GL(V^{\tn n})$, where $  \GL(V^{\tn n})$ is the group of invertible linear maps $V^{\tn n} \to V^{\tn n}$. If $\g$ is semisimple, ${\langle -, -\rangle}$ is the Cartan-Killing form and $r$ is the infinitesimal R-matrix associated to ${\langle -, -\rangle}$, then this representation of the braid group with $n$-strands is equivalent to the representation of the braid group derived from the $R$-matrix of the quantum group $U_q(\g)$, for $q=e^{h}$, a beautiful fact known as Kohno's {Theorem}, \cite{Kh1}; see also \cite{Dr,Ka}.

The holonomy of the  KZ-connection cannot immediately  be extended to links in $S^3$. This is because the forms $\w_{ab}$ explode at minimal and maximal points, when two particle trajectories $z_i\colon I \to \C$ and $z_{i+1}\colon I \to \C$ collide. Nevertheless, the KZ-connection holonomy can be regularized at maximal and minimal points, as a power series in $h$ \cite{AF1,LM}. After adding an anomaly  correction term, this leads to knot invariants \cite{AF1,BN,Ko,LM}, coinciding with the usual quantum group knot invariants \cite{Dr,Ka}.

For a positive integer $n$, we can also consider the Lie algebra $\ch_n$, formally generated by the symbols $r_{ab}$, where $1\leq a<b\leq n$, satisfying the infinitesimal braid group relations as in \eqref{ibr1} and \eqref{ibr2}. Call it the {\it Lie algebra of horizontal chord diagrams (in the 1-manifold consisting of $n$ parallel strands)}. Consider the connection form $A=\sum_{1 \leq a<b \leq n} \w_{ab}r_{ab}$ taking values in $\ch_n$. By using Chen integrals \cite{Ch}, as in \cite{Kh1,Ko,BN,Ka}, we can define the holonomy of this connection, living in the space of formal power series over the universal enveloping algebra $\U(\ch_n)$ of $\ch_n$. As before this holonomy can be regularized at maximal and minimal points of embedded links \cite{AF1,LM}, defining a knot invariant with values in the space of formal power series in the Hopf algebra of chord diagrams in the circle. This invariant is called the Kontsevich integral, and can be proven to be a universal Vassiliev invariants of knots; \cite{Ka,Ko,BN}..

In this article we present a categorification of the Lie algebra $\ch_n$ of horizontal chord diagrams. We do not address the seemingly  related categorification of the important case of the Hopf algebra of chord diagrams in the circle (using the framework of this article), which we intend to postpone to a future publication.

The context we will use to categorify  $\ch_n$  is the context of categorical group 2-connections on a manifold $M$ \cite{BS1,BrMe,MP,GP}. It is well known \cite{BS} that a Lie categorical group can be equivalently described by a crossed module $\G=(\d\colon H \to G,\t)$; see also \cite{BL}. Here $\d\colon H \to G$ is a Lie group morphism and $\act$ is a left action of $G$ on $H$ by automorphisms. The Lie algebras of these can be arranged into a differential crossed module $\mathfrak{G}=(\d\colon \h \to \g,\t)$. For details see \cite{B,BL,BC}, and also \cite{FMP1}. Locally a {(fake-flat)} 2-connection looks like a pair $(A,B)$, where $A$ is a 1-form in $M$ with values in $\g$ and $B$ is a 2-form in $M$ with values in $\h$, such that $\d(B)=dA+\frac{1}{2} A \wedge A$, the curvature of $A$. A 2-connection is said to be flat if the {curvature} 3-form $dB+A\wedge^\t B$ vanishes.

As principal $G$-bundles over $M$ with connection have a $G$-valued  holonomy assigned to closed paths $\gamma \colon [0,1]=D^1 \to M$,  2-bundles with a 2-connection taking values in {$\mathfrak{G}=(\d\colon \h \to \g,\t)$} have a 2-dimensional holonomy, assigned to maps $\Gamma \colon D^2=[0,1]^2 \to M$, and taking values in $H$, as well as an underlying 1-dimensional holonomy assigned to paths, living in $G$. For details see \cite{BS1,BaH,SW1,SW2,FMP1,FMP2,FMP3}.

This 2-dimensional holonomy of a 2-connection is invariant under homotopy of maps $\Gamma\colon [0,1]^2 \to M$, stable in the boundary of the square, and factoring through a 2-dimensional submanifold, a consequence of the invariance of the 2-dimensional holonomy under thin homotopy \cite{BS1,SW1,FMP1,FMP2}. If the underlying 2-connection is flat, then the 2-dimensional holonomy depends only on the homotopy class (relative to the boundary) of the map $\Gamma \colon [0,1]^2 \to M$, a fact which we will explore in this article in the context of {braided surface}s.

\begin{figure} 
{\centerline{\relabelbox 
\epsfysize 4cm
\epsfbox{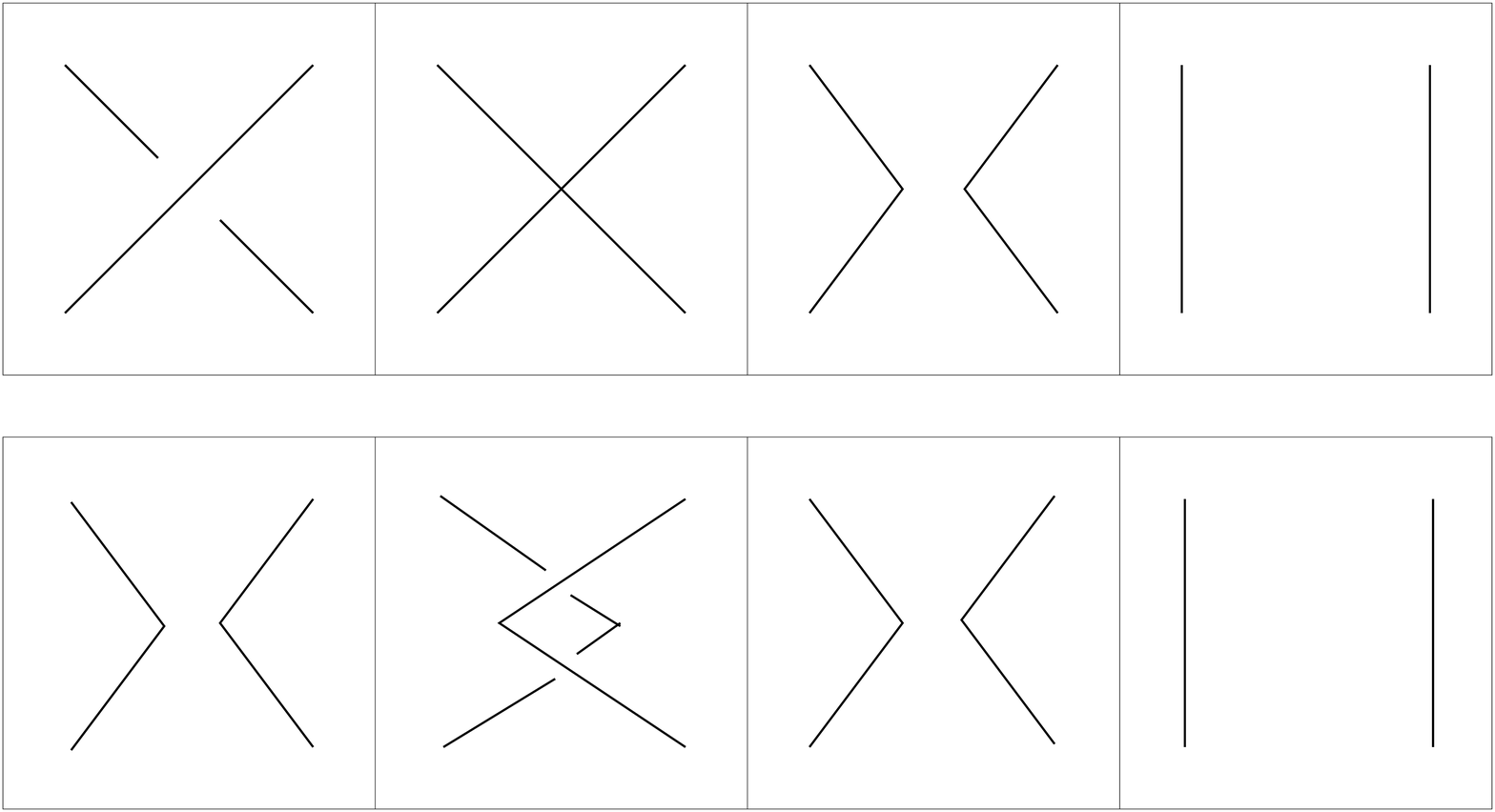}
\endrelabelbox}}
\caption{{\label{sbraids} {Braided surfaces} with branching numbers $1$ and $0$. {In the first case, two strands move to meet each-other and then recombine the other way, in the second case we simply have a Reidemeister-II move, followed by its inverse}. }}
\end{figure}

{Let $s$ be a non-negative integer. A (simple) {braided surface}  $b_1 \ra{\S} b_2$ \cite{CKS} (called a braid cobordism in \cite{KT}), of branching number  $s$, connecting the braids $b_1$ and $b_2$, seen as embedded 1-manifolds in $D^3=[0,1]^3$, is an embedded 2-manifold $\S$ in $[0,1]^4=[0,1]^3 \times [0,1]$, defining an embedded cobordism between $b_1$ and $b_2$. We further suppose that the projection of $\S$ onto $\{(0,0)\} \times D^2$ is a simple branched cover with $s$ branching points, and moreover that the intersection  of $\S$ with $[0,1]^2 \times \{\pm 1\} \times [0,1]$ does not depend on the last variable. See figure \ref{sbraids} for two examples of {braided surface}s, described by their intersections with $D^3 \times \{t\}$, with $t\in [0,1]$.}

Any {braided surface} $b_1 \ra{\S} b_2$ with {branching number} $0$ defines a map $\S'\colon D^2 \to \C(n)/S_n$, restricting to $b_1\colon D^1 \to \C(n)/S_n$ and to $b_2\colon D^1 \to \C(n)/S_n$ on the top and bottom of $D^2$, with $\S'$ being constant on the left and right sides of $D^2$; see \cite[1.6]{CKS}. {If $\S$ has {branching number} $s$ then $\S'$ is defined on $D^2$ minus a set with $s$ points, on which $\S'$ has a very particular type of singularities, see subsection \ref{cssb}.}

We therefore aim to define flat 2-connections in the configuration spaces $\C(n)$ and $\C(n)/S_n$, since these naturally assign a 2-dimensional holonomy to a {braided surface} with  no branch points.  Namely, choose  a flat 2-connection $(A,B)$ in a (for the sake of simplicity trivial) 2-bundle over $\C(n)$.
 Then if we have a {braided surface} $b_1 \ra{\S} b_2$, without branch points, connecting the braids $b_1$ and $b_2$,  there will exist one-dimensional holonomies $H(b_1)$ and $H(b_2)$ of $b_1$ and $b_2$ (which will now not necessarily be invariant under braid isotopy), {related by the 2-dimensional holonomy $H(\S')$ of $\S$: 2-categorically we have a 2-morphism $H(b_1) \ra{H(\S')} H(b_2)$.} This {two-dimensional} holonomy will be invariant under {braided surface} isotopy, since the 2-connection has vanishing {curvature}. Moreover it is functorial with respect to the two obvious, horizontal and vertical, compositions of braided surfaces (without branch points). An open problem is whether this {two-dimensional} holonomy can be regularized in the case when $\S$ has branch points and therefore the map $\S'\colon D^2 \to \C(n)/S_n$ has (a very particular type of) singularities.

As Lie algebras act on vector spaces, differential crossed modules (categorically) act on chain complexes of vectors spaces, see subsection \ref{seccatreps} (we will carefully address this kind of categorical representations). This is because given a chain complex $\V$ of vector spaces we can define a differential crossed module $\gl(\V)$ of {chain map}s $\V \to \V$ and homotopies (up to 2-fold homotopies) of $\V$, see subsection \ref{dcmcc}. This appeared  in \cite{FMM}, borrowing ideas from \cite{KP,BL,BC,FB}. Categorical representations of crossed modules are also treated in \cite{BM,E}.  Defined like this, categorical representations of differential crossed modules have a natural tensor product, acting in the usual tensor product $\btn$ of chain complexes.   

Given a chain complex $\V$, a positive integer $n$, and chain maps $r_{ab}\colon \V^{\btn n} \to \V^{\btn n}$, where $1\leq a<b\leq n $, as well as chain homotopies (up to 2-fold homotopy) $K_{abc}$ and $K_{bac}$, where $1\leq a<b<c \leq n$, of $\V^{\btn n}$,  we find necessary and sufficient conditions for a local 2-connection $(A,B)$ in $\C(n)$ of the form 
\begin{align}
 A&=\sum_{a <b} \w_{ab} r_{ab}\\
  B &= \sum_{a<b<c} K_{bac} \, \w_{ab}\wedge\w_{ac} + K_{abc} \, \w_{ab}\wedge\w_{bc}
\end{align}
to be flat, see Theorem \ref{iffflat}. We also give sufficient conditions for the 2-dimensional holonomy of it to descend to a 2-dimensional holonomy in $\C(n)/S_n$, namely so that the pair $(A,B)$ is invariant under the action of the symmetric group (this is called the totally symmetric case). This is contained in Theorem \ref{thm13}. 

The relations we get for the chain maps $r_{ab}$ as well as the homotopies   $K_{abc}$ and $K_{bac}$ lead to the definition of the differential crossed module of totally symmetric horizontal 2-chord diagrams $2\mathfrak{ch}_n=(\d\colon 2\ch_n \to \ch_n^+)$. This differential crossed module is defined by generators and relations in Section \ref{dcm}, Theorem \ref{refnow}, as the quotient of a free differential crossed module.

Given a differential {crossed module} $(\d\colon \h \to \g)$ acting on a chain complex $\V$, one would like to find the conditions that the tensors $r\in \g \tn \g$ and $P \in \bar{\mathfrak{U}}^{(3)}$, which is a quotient of $\g \tn \g \tn \h \oplus \g \tn \h \tn \g \oplus \h\tn \g \tn \g$, provided with a natural map $\hat{\d}$ onto $\g \tn \g \tn \g$, should satisfy in order that, by considering the associated chain maps $\bar{\phi}_{ab}(r)$ and chain homotopies $\bar{\phi}_{abc}(P)$ in $\V^{\btn n}$, the 2-connection $(A,B)$ with 
\begin{align}
A &= \sum_{a<b} \w_{ab} \, \bar{\phi}_{ab}(r) \, .\\
B &= \sum_{a<b<c} \w_{ab}\wedge\w_{ac} \, \bar{\phi}_{bac}(P) + \w_{ab}\wedge \w_{bc} \, \bar{\phi}_{abc}(P)
\end{align}
is flat and totally symmetric. These conditions are below and define what we call a {\it totally symmetric infinitesimal {2-R}-matrix}:
\begin{equation}
\begin{split}
& r_{12}=r_{21} \\
&\hat{\d}(P)=[r_{12}+r_{13},r_{23}]\\
& r_{14}\act (P_{213} + P_{234}) + (r_{12} + r_{23} + r_{24}) \act P_{314} - (r_{13}+r_{34})\act P_{214} = 0 \\
& r_{23}\act (P_{214} + P_{314}) -r_{14}\act (P_{423} +P_{123}) = 0 \\
& P_{123}+P_{231}+P_{312}=0 \\ 
& P_{123}=P_{132}
\end{split}
\end{equation}
All of this is explained in Section \ref{defir}.


\section*{Open Problems}
Several open problems come out of this article. 

First of all, it is well known \cite{Br} that crossed modules of groups $(\d\colon H \to G)$ are classified, up to weak equivalence, by group cohomology classes $k^3$ in $H^3(\coker(\d), \ker(\d))$, a result that appeared originally in \cite{ML}. Similarly \cite{G2} differential crossed modules are classified, up to weak equivalence (or what is the same by equivalence in the larger category of Lie 2-algebras \cite{BL,BC}), by a Lie algebra cohomology class $k^3 \in H^3(\mathfrak{k},M)$. Here a differential crossed module $\d\colon \h \to \g$ sits inside the exact sequence of Lie algebras 
$$\{0\} \to M \to \h \ra{\d} \g \ra{{\rm proj}} \mathfrak{k}\to \{0\} ,$$
with $M$ abelian, and $\mathfrak{k}$ has an obvious induced action on $M$, well defined by the differential crossed module axioms. 
Given a differential crossed module $(\h\to \g)$, the associated cohomology class (the $k$-invariant) is denoted by $k^3(\h\to \g)$, and we say that $(\h \to \g)$ geometrically realizes $k^3$.

\begin{Problem}
Describe the kernel $M_n$ of the boundary map $\d\colon 2\ch_n \to \ch_n^+$ in the differential crossed module $2\mathfrak{ch}_n=(\d\colon 2\ch_n \to \ch_n^+)$ of totally symmetric horizontal 2-chord diagrams. (The cokernel is the Lie algebra $\ch_n$ of horizontal chord diagrams,  generated by $r_{ab}$, where $1\leq a<b \leq n$, subject to the infinitesimal braid relations \eqref{ibr1} and \eqref{ibr2}.) Address whether the associated cohomology class $k^3(2 \mathfrak{ch}_n) \in H^3(\ch_n,M_n)$ is trivial or not.
\end{Problem}
Any simple Lie algebra $\k$ comes \cite{BSCS} with a cohomology class $k^3\in H^3(\k,\C)$, namely $k(X,Y,Z)=<X,[Y,Z]>$, where ${\langle -, -\rangle}$ is the Cartan-Killing form. Explicit constructions (defined up to weak equivalence) of  differential crossed module geometrically realizing this cohomology class appear in \cite{BSCS,Wa}, {leading to the definition of the String Lie-2-algebra.}

\begin{Problem} Given a simple Lie algebra $\k$, address whether there exist totally symmetric infinitesimal {2-R}-matrices $(r,P)$ in the crossed modules associated to the  cohomology class $k^3\in H^3(\k,\C)$. It is important that the projection map ${\rm proj}\colon \g \to \k$ maps $r$ to the infinitesimal R-matrix in $\k$ coming from the Cartan-Killing form in $\k$, so that we would be obtaining a categorification of the braid group representation coming from the quantum group $\U_q(\k)$.
\end{Problem}

By considering Chen integrals as in \cite{BN,Ka}, we can define given a {braided surface} $b_1 \ra{\S} b_2$, without branch points, with associated map $\S'\colon D^2 \to \C(n)/S_n$, a holonomy $H(b_1) \ra{H(\S')} H(b_2)$, where $H(b_1)$ and $H(b_2)$ take values in the algebra of formal power series in the universal enveloping algebra $\U(\ch^+_n)$, a Hopf algebra, and $H(\S')$ takes values in the algebra of formal power series in $\U(2\ch_n)$.
\begin{Problem}
Extend this holonomy to the case when $\S$ has branch points. This will require some form of regularization since, in the general case, the associated map $\S'\colon D^2 \setminus \{\textrm{branch points}\} \to \C(n)/S_n$ will not be defined in all of $D^2$, see subsection \ref{cssb}, however having a very particular type of singularities. It would be very important to analyze whether the first {braided surface} of figure \ref{sbraids} has a non-trivial 2-dimensional holonomy or not.
\end{Problem}

\begin{Problem}
Is it possible to define a Hopf algebra crossed module of 2-chord diagrams in the 2-sphere from the relations defining $\mathfrak{ch}_n$?
\end{Problem}

\begin{Problem}
As infinitesimal R-matrices  in a Lie algebra come naturally from  invariant non-degenerate symmetric bilinear forms, it would be important to find a simple geometric way to construct infinitesimal {2-R}-matrices.
\end{Problem}

\section{Differential crossed modules}

\subsection{Crossed modules of Lie groups and algebras}

For details on (Lie) crossed modules see, for example, \cite{B1,BL,B,BHS,FM1,FMP1}, and references therein.

\begin{Definition}[Lie crossed module]\label{LCM}
 A crossed module ${\G= ( \d\colon H \to  G,\t)}$ is given by a group morphism $\d\colon H \to G$ together with a left action $\t$ of $G$ on $H$ by automorphisms, {such that}:
\begin{enumerate}
 \item $\d(g \t h)=g \d(h)g^{-1}; \fo g \in G \an  h \in H,$
  \item $\d(h) \t h'=hh'h^{-1};\fo  h,h' \in H.$
\end{enumerate}
If both $G$ and $H$ are Lie groups, $\d\colon H \to G$ is a smooth morphism,  and the left action of $G$ on $H$ is smooth then $\G$ will be called a Lie crossed module. A pre-crossed module is defined analogously, however skipping  the second condition.
\end{Definition}

A morphism $\G \to \G'$ from the crossed module ${\G= ( \d\colon H \to  G,\t)}$  to the crossed module {$\G'=(\d'\colon H' \to G',\t')$} is given by a pair of maps $\f\colon G \to G'$ and  $\psi\colon H \to H'$ which make the following diagram commutative,
$$ \begin{CD}
  H @>\d>> G \\
 @V \psi VV   @VV \f V \\
  H' @>\d'>> G' \\
   \end{CD}
$$ and such that $\psi(g \t e)=\f(g) \t' \psi(e)$ for each $ e \in H$ and each $ g \in G$.

\begin{Example}
 Let $G$ be a Lie group and $V$ a vector space carrying a representation $\rho$ of $G$. Then we can define a crossed module $(V \ra{v \mapsto 1_G} G,\rho)$. 
\end{Example}
\begin{Example}
 Let $G$ be a connected Lie group and $\Aut(G)$ be the Lie group of all automorphisms of $G$. We have a left action of $\Aut(G)$ on   $G$ by automorphisms, where $f \t g=f(g)$, for $f \in \Aut(G)$ and $g \in G$. Together with the  map $g \in G \mapsto \Ad_g$ which sends $g \in G$ to the automorphism $h \mapsto ghg^{-1}$ this defines a crossed module. 
\end{Example}

Given a Lie crossed module ${\G= ( \d\colon H \to  G,\t)}$, we have an induced Lie algebra map $\d\colon \e \to \g$, and a  derived action of $\g$ on $\e$ (also denoted by $\t$). This forms  a differential crossed module, in the sense of the following definition - see \cite{B1,BC,BS1,FMP1,FMP2}.

\begin{Definition}[Differential crossed module] A differential crossed module ${\mathfrak{G}=(\d \colon \e \to  \g,\t )}$ is given by a Lie algebra morphism $\d\colon \e \to \g$ together with a left action of $\g$ on the underlying vector space of  $\e$, {such that}:
\begin{enumerate}
 \item For any $X \in \g$ the map $\xi \in \e \mapsto X \t \xi \in \e$ is a derivation of $\e$, in other words 
\begin{equation}
X \t [\xi,\nu]=[X \t \xi,\nu]+[\xi, X \t \nu];\fo X \in \g, \ane \xi ,\nu\in \e.
\end{equation}
\item The map $\g \to \Der(\e)$ from $\g$ into the derivation algebra of $\e$ induced by the action of $\g$ on $\e$ is a Lie algebra morphism, in other words:
\begin{equation}
[X,Y] \t \xi=X \t (Y \t \xi)-Y \t(X \t \xi); \fo X,Y \in \g \an \xi \in \e \,,
\end{equation}
\item 
\begin{equation}
\d( X \t \xi)= [X,\d(\xi)];\fo X \in \g, \ane \xi\in \e\,,\label{fid} \end{equation}
\item \begin{equation} \d(\xi) \t \nu=[\xi,\nu];  \fo \xi,\nu\in \e\,.\label{sid}\end{equation}
\end{enumerate}
As before, a differential  pre-crossed module is defined analogously, but skipping  the fourth condition.
\end{Definition}
In any differential crossed module we have:
\begin{equation}
\d(\xi) \t \nu=[\xi,\nu]=-[\nu,\xi]=\d(\nu)\t \xi, \fo \xi,\nu \in \e.  
\end{equation}

We have a functor which sends a Lie crossed module to its associated differential crossed module. On the other hand, given a  differential crossed module ${\mathfrak{G}=(\d \colon \e \to  \g,\t )}$ there exists a unique (up to isomorphism) crossed module of simply connected Lie groups ${\G=(\d \colon H \to  G,\t )}$ whose differential form is $\mathfrak{G}$.


\subsection{Differential crossed modules from complexes of vector spaces}\label{dcmcc}
\subsubsection{Short complexes}\label{scomp}
  
Let $\V=(V \ra{\d} U)$ be a short complex of (finite dimensional) vector spaces. In other words $V$ and $U$ are vector spaces and $\d\colon V \to U$ is a linear map. Let us define a differential crossed module $\gl(\V)=\big ( \beta\colon \gl^1(\V) \to \gl^0(\V),\t \big )$. 
This is a well known construction; see for example \cite{BC,BL,FB}. For details on the construction of the  associated Lie crossed module  $\GL(\V)=\big ( \beta\colon \GL^1(\V) \to \GL^0(\V),\t \big )$ see \cite{FMM}.

Consider the algebra  $\Hom^0(\V)$  chain maps $f\colon \V \to\V$, with composition as product. The  Lie algebra  $\gl^0(\V)$ is identical to $\Hom^0(\V)$ as a vector space,  with bracket given by the commutator in $\Hom^0(\V)$; in other words $[F,F']=F\circ F'-F' \circ F$, for any two chain maps $F,F'\colon \V \to \V$.

The Lie algebra $\gl^1(\V)$ is given by the vector space $\Hom^1(\V)$ of all maps $s\colon U \to V$, with bracket given by 
$$[s,t]=s\d t-t\d s \, .$$ 
The map $\b \colon \gl^1(\V) \to \gl^0(\V)$ such that
$$\b(s)=(s \d,\d s) $$ is a morphism of Lie algebras. A left action of $\gl^0(\V)$ on $\gl^1(\V)$, by derivations, can be defined as
$$(f_V,f_U) \t s=f_V s- sf_U.$$ 
Simple calculations prove that this indeed defines a differential crossed module.


\subsubsection{Long complexes}\label{lcomp}

By slightly modifying the previous construction in Section \ref{scomp}, we can construct a  differential crossed module $$\gl(\V)=\left (\b\colon \gl^1(\V) \to \gl^0(\V) ,\t \right)$$ from any complex of vector spaces $\V=(\dots \ra{\d} V_n \ra{\d} V_{n-1} \ra{\d}\dots)$; this appeared in \cite{FMM,FMP3}, with ideas borrowed from \cite{KP}. {We can also analogously construct  an associated Lie crossed module  $\GL(\V)=\big ( \beta\colon \GL^1(\V) \to \GL^0(\V),\t \big )$; see \cite{FMM}.}

First of all define a Lie algebra $\gl^0(V)$, given by all chain maps $f\colon \V \to \V$, with the usual commutator of {chain map}s giving the Lie algebra structure.

A degree $n$ map $h\colon \V \to \V$ is given by a sequence of linear maps $h_i\colon V_i \to V_{i+n}$, without any {compatibility} relations with $\d$. We denote the vector space of degree-$n$ maps by $\Hom^n(\V)$.  We can define a Lie algebra structure on the vector space $\Hom^1(\V)$ of degree 1 maps where:
$$[s,t]=s\d t -t \d s +st\d-ts \d.$$
The bilinearity and antisymmetry of this bracket are immediate, whereas Jacobi identity follows from an explicit calculation. Moreover, the usual chain-complex boundary map $\beta\colon \Hom^1(\V) \to \gl^0(\V)$ such that 
$$\b(s)=\d s+s \d $$
is a Lie algebra morphism. 
There exists an action of $\gl^0(\V)$ on $\Hom^1(\V)$ such that:
$$f \t s=fs-sf .$$
An explicit calculation shows that this in  an action by derivations. Moreover we have
$$ \beta(f \t s)=[f, \beta(s)], \fo f \in \gl^0(\V) \an s \in \Hom^1(\V).$$

We do not always have a differential crossed module since the second Peiffer identity $[s,t]=\b(s) \t t$ may fail in general, unless we are considering a complex of length two. 
Consider the  map $\beta'\colon \Hom^2(\V) \to \Hom^1(\V)$ such that $$\b'(h)=h\d -\d h.$$ Then $\b'(\Hom^2(\V))$ is a $\gl^0(\V)$-invariant Lie algebra ideal of $\Hom^1(\V)$,  contained in $\ker(\b)$. In fact for $h \in \Hom^2(\V)$ and $f \in \gl^0(\V)$ we have:
$$f \t \b'(h)=\b'(f h-h f), $$
and also 
$$ [s,\b'(h)]=\b'(h\d s-\d h s) .$$

We can  therefore define a Lie algebra $$\gl^1(\V)=\frac{\Hom^1(\V)}{\b'(\Hom^2(\V))},$$ 
provided with a (quotient) map $$\b\colon \gl^1(\V) \to \gl^0(\V)$$
and a quotient action $\t$ by derivations of $\gl^0(\V)$ on $\gl^1(\V)$. To prove this is a crossed module of Lie algebras we must check $\b(s)\t t =[s,t]$, in the quotient. This follows from  
\begin{align}
 \b(s) \t t -[s,t]=\b'(st), \fo s,t \in \Hom^1(\V).
\end{align}


\section{{A 2-connection categorifying the Knizhnik-Zamolodchikov connection and the differential crossed module of  (totally symmetric) horizontal $2$-chord diagrams}}

\subsection{Local 2-connections} 
\label{2conn}
Fix a manifold $M$. Given a vector space ${U}$, we denote the vector space of ${U}$-valued  differential $n$-forms in $M$ as $\Omega^n(M,U)$. Let $V$ and $W$ be vector spaces. Suppose we have a bilinear map $ {L}\colon  U\times V \to W$. If we are given ${U}$ and $V$ valued forms $\mu\in \Omega^a (M,U)$ and $\nu \in \Omega^b(M,V)$ we define  the $W$-valued $(a+b)$-form  $\mu \wedge^{{L}}\nu$ in $M$  as:
$$\mu \wedge^{{L}} \nu= \frac{(a+b)!}{a!b!}{\rm Alt}(\mu \otimes^{{L}} \nu) \in \Omega^{a+b}(M,W).$$
Here $\mu\otimes^{{L}} \nu$ is the covariant tensor ${{L}}\circ (\mu \times \nu)$ and ${\rm Alt}$ denotes the natural projection from the vector space of  $W$-valued  covariant tensor fields in $M$ onto the vector space of $W$-valued differential forms in $M$.

Given a Lie crossed module $\G=(\d\colon H \to G,\t)$ with associated differential crossed module $\lX=(\d\colon \h \to \g,\t)$, a $\lX$-valued {(and fake-flat)} local 2-connection pair $(A,B)$ in $M$ is given by a $\g$-valued  1-form $A\in \Omega^1(M,\g)$ and a $\h$-valued 2-form $B\in \Omega^2(M,\le)$ such that 
\begin{equation}
\label{db}
\d(B)=\F_A\doteq d A+\frac{1}{2}A\wedge^{{[-,-]}} A \, .
\end{equation}
{(Here the bilinear map used to define the exterior product is given by the Lie bracket $[-,-]$).} This means that for  vector fields $X$ and $Y$ in $M$ we have:
\begin{equation}
\d(B(X,Y))=d A(X,Y)+[A(X),A(Y)].
\end{equation}
{Note that $\F_A=d A+\frac{1}{2}A\wedge^{{[-,-]}} A$ is the usual curvature 2-form of the connection form $A$.}
The {curvature} 3-form of a local 2-connection pair $(A,B)$ is given by
\begin{equation}
\label{2curv}
\M_{(A,B)}=d B+A\wedge^\t B. 
\end{equation}
{(In this case the bilinear map appearing in the exterior product is $(X,v)\in \g\times \le \mapsto X \t v \in \le$.) For any vector fields $X,Y$ and $Z$ in $M$ we therefore must have:}
\begin{equation}
\M_{(A,B)}(X,Y,Z) = d B (X,Y,Z)+A(X)\t B(Y,Z)+A(Y)\t B(Z,X)+A(Z)\t B(X,Y).
\end{equation}
A local 2-connection is said to be flat if its {curvature} 3-form vanishes.


\subsection{The 2-dimensional holonomy of a local 2-connection}

Consider a Lie crossed module $\G=(\d\colon H \to G,\t)$ with associated differential crossed module $\lX=(\d\colon \h \to \g,\t)$.
A local 2-connection in a manifold  determines a 2-dimensional holonomy, in the sense we now present.

\subsubsection{Paths and 2-paths in $M$}

A path is by definition a piecewise smooth map $\gamma\colon D^1=[0,1] \to M$. Paths $\gamma,\gamma'$ in $M$ can be concatenated to give a path $\gamma \gamma'$ in $M$ if the end-point $\d_1^+(\gamma)\doteq \gamma(1)$ of $\gamma$ coincides with the initial point $\gamma'(0)=\d_1^-(\gamma')$ of $\gamma'$.
As usual
$$\gamma\gamma'(s)=\left \{ \begin{CD}
                             \gamma(2s), s \in [0,1/2]\\
                             \gamma'(2s-1), s \in [1/2,1] 
                            \end{CD} \right.
 $$

A 2-path is by definition a map $\Gamma\colon D^2=[0,1]^2 \to  M$, piecewise smooth for some paving of the square $D^2$ by polygons. We also assume that $\d^+_1(\Gamma)=\Gamma(1,s)$ and $\d^-_1(\Gamma)=\Gamma(0,s)$ {are each} constant paths. Define also (not necessarily constant) $2$-paths $\d^\pm_2(\Gamma)$ as being the restrictions $\Gamma(t,1)$ and $\Gamma(t,0)$ of $\Gamma$. 

Note that we have horizontal and vertical concatenations of $2$-paths $\Gamma$ and $\Gamma'$, defined as long as they coincide on the relevant side of the square.


\subsubsection{Edges and {disk}s in a crossed module $\G$.}
 {Let $\G=(\d\colon H \to G,\t)$ be a crossed module.
An edge in $\G$ is by definition an arrow colored with an element $g \in G$, in other words a diagram of the form:}
$${* \ra{g} *, \textrm{ where } g \in G.} $$
Edges in $\G$ can be composed in the obvious way:
$$* \ra{g} * \ra{g'} *=  * \ra{gg'} *. $$
Analogously, {disk}s in $\G$ are {diagrams of the form:}
\begin{equation}
 \xymatrix{ &\ast\ar@/^1pc/[rr]^{g'}\ar@/_1pc/[rr]_{g} & e  &\ast}
\end{equation}
where $g,g' \in G$ and $e \in H$ is such that $\d(e)^{-1}g=g'$. {Disks in $\G$ can be composed horizontally and vertically. The horizontal composition of disks in $\G$ is always defined for any two disks and has the form:}
$$  
\hspace{-1.7cm}
\xymatrix{ &\ast\ar@/^1pc/[rr]^{g_1'}\ar@/_1pc/[rr]_{g_1} & e   &\ast\ar@/^1pc/[rr]^{g_2'}\ar@/_1pc/[rr]_{g_2} & e'  &\ast &=& \ast\ar@/^1pc/[rr]^{g_1'g_2'}\ar@/_1pc/[rr]_{g_1g_2} & (g_1 \t e')e  &\ast} .
$$
{The vertical composition of two disks in $\G$ is only well defined if the edge in $\G$ assigned to the bottom of the first disk coincides with the edge assigned to the top of the second disk, and it has the form:}
$$ \xymatrix{ &\ast\ar@/^1pc/[rr]^{g''}\ar@/_1pc/[rr]_{g'} & e'   &\ast \\ &&&&=&\ast\ar@/^1pc/[rr]^{g''}\ar@/_1pc/[rr]_{g} & ee'  &\ast 
\\ & \ast \ar@/^1pc/[rr]^{g'}\ar@/_1pc/[rr]_{g} & e  &\ast  }$$
These horizontal and vertical compositions of {disk}s in $\G$ {are associative and further they} satisfy the interchange condition \cite{BHS}, familiar in two-dimensional category theory.

\subsubsection{The form of a 2-dimensional holonomy}

Let $M$ be a manifold, and let $G$ be a Lie group with Lie algebra $\g$. Let $\gamma\colon [0,1] \to M$ be a piecewise smooth map. Let $A \in \Omega^1(M,\g)$ be a $\g$-valued 1-form in $M$. We can integrate $A$ with respect to $\gamma$ in the usual way, by defining ${\stackrel{A}{g}}\gamma(t) \, \in G$ as the solution of the differential equation in $G$:
$$ \frac{d}{dt} {\stackrel{A}{g}}_\gamma(t)={\stackrel{A}{g}}_\gamma(t) \,\, A\left( \frac{d}{d t} \gamma(t)\right) ,$$
with initial condition  ${\stackrel{A}{g}}_\gamma(0)=1_G$. Put ${\stackrel{A}{g}}_\gamma\doteq {\stackrel{A}{g}}_\gamma(1)$.
 If $\gamma_1$ and $\gamma_2$ are piecewise smooth maps with $\gamma_1(1)=\gamma_2(0)$, we have that ${\stackrel{A}{g}}_{\gamma_1\gamma_2}={\stackrel{A}{g}}_{\gamma_1}{\stackrel{A}{g}}_{\gamma_2}$.

Let $\G=(\d\colon H\to  G,\triangleright)$ be a Lie crossed module and let  $\mathfrak{G}=(\d\colon \le\to \g,\t)$ be the associated differential crossed module. If we have $B \in  \Omega^2(M,\le)$ with $\d(B)=F_A\doteq d A+\frac{1}{2} A \wedge^{[\, , \, ]} A$, {which therefore means that} $(A,B)$ is a local 2-connection, we define 
$\stackrel{(A,B)}{e_\Gamma}(t,s) \, \in H$ as being the solution of the differential equation (where we put $\gamma_s(t)=\Gamma(t,s)$)
$$\frac{\d}{\d s} \stackrel{(A,B)}{e_\Gamma}(t,s)=\stackrel{(A,B)}{e_\Gamma}(t,s)  \int_0^t \stackrel{A}{g}_{\gamma_{s}(t')} \t B \left(\frac{\d}{\d t'}\gamma_{s}(t'),\frac{\d}{\d s}\gamma_{s}(t')\right)dt' $$ 
with initial conditions $$ \stackrel{(A,B)}{e_\Gamma}(t,0)=1_H \, , \quad \forall t \in [0,1].$$

Put $\stackrel{(A,B)}{e_\Gamma}=\stackrel{(A,B)}{e_\Gamma}(1,1)$. The following result is proven in \cite{BS1,SW2,FMP1,FMP2,FMP3}.

\begin{Theorem}
Let $M$ be a smooth manifold with a local 2-connection pair $(A,B)$, taking values in the differential crossed module  $\lX=(\d\colon \h \to \g,\t)$, associated to the Lie crossed module  $\G=(\d\colon H \to G,\t)$. The assignment $\Gamma \mapsto {\rm Hol}(\Gamma)$,  which {to} a 2-path $\Gamma$ associates
$${\rm Hol}(\Gamma)=  \xymatrix{ &\ast\ar@/^2pc/[rr]^{g^A_{\d^{+}_{2}(\Gamma)}}\ar@/_2pc/[rr]_{g^A_{{\d^{-}_{2}(\Gamma)}}} & e_\Gamma^{(A,B)}  &\ast}$$
preserves horizontal and vertical composites. In other words
$${\rm Hol}({\Gamma \Gamma'})={\rm Hol}({\Gamma}) {\rm Hol}({\Gamma'}) $$
and 
$${\rm Hol} \left( \substack{\Gamma \\ \Gamma'}\right)=\substack{{\rm Hol}(\Gamma) \\ {\rm Hol}(\Gamma')} \, .$$
\end{Theorem}

As is the case of 1-dimensional holonomy, the variation of the holonomy when we vary the 2-paths is ruled by the {curvature 3-form} \cite{BS1,SW1,SW2}. It is proven in \cite{FMP1,FMP2} that:
\begin{Theorem}
Suppose $(A,B)$ is flat and that $\Gamma$ and $\Gamma'$ are homotopic, relative to the boundary of $D^2$. Then ${\rm Hol}(\Gamma)={\rm Hol}(\Gamma')$.
\end{Theorem}


\subsection{Configuration spaces and braided surfaces}\label{cssb}

\begin{Definition} {Let $n$ be a positive integer}.
 The configuration space $\C(n)$ of $n$ distinguishable particles in the complex plane $\C$ is the set of tuples $(x_1,\dots, x_n)$ in $\C^n$ such that $x_i \neq x_j$ if $i\neq j$. This space has an obvious properly discontinuous action of the symmetric group $S_n$ by permutation of coordinates. We thus define the space of $n$ {indistinguishable} particles as $\C(n)/S_n$, a manifold of dimension $2n$.
\end{Definition}

The pure braid group $P_n$ is isomorphic to the fundamental group of $\C(n)$, whereas the braid group $B_n$ is isomorphic to the fundamental group of $\C(n)/S_n$. 

A particular set of maps we would like to consider are the branching maps $m^\pm\colon B^2 \to \C(2)/S_2$ from the unit ball of $\C$ (minus the origin) to the configuration space $\C(2)/S_2$, defined as (in polar coordinates): 
$$m^+(\theta,r)=\big(-\exp(i \theta /2),\exp(i\theta /2)\big)r$$
and 
$$m^-(\theta,r)=\big(-\exp(-i\theta /2),\exp(-i\theta /2)\big)r .$$
These correspond to the type of catastrophe  that happens when we try to interpret the  first {braided surface} of figure \ref{sbraids} as a map $D^2 \to \C(2)$.
If we restrict to the boundary $S^1$ of the 2-ball we thus obtain the standard generators of the braid group $B_2$. These branching maps can be generalized to maps $m^{\pm}_i\colon B^2 \to \C(n)/S_n$ creating a branch point connecting the $i$-th and $(i+1)$-strands of a braided surface.

The following is a slightly non-standard definition of  braided surfaces, however well adapted to address their {two-dimensional} holonomy. {For a detailed description of the concept of a braided surface we refer for example to \cite{CKS}.}

\begin{Definition}[Braided surface]
{Let $s$ be a non-negative integer. A  braided surface $\S$ (of branching number $s$, and degree $n$) is given by a map $\S'\colon D^2 \setminus \sigma(\S) \to \C(n)/S_n$ (where as usual $D^2=[0,1]^2$), such that:}
\begin{enumerate}
\item {The set $\sigma(\S)$ is a set with $s$ points, contained {in} the interior of $D^2$.  Each element of $\s(\S)$ corresponds therefore to some branch point of $\S$.}
 \item The map $\S'$ is smooth. 
 \item There exists a disk around each element of $ \sigma(\S)$ where $\S'$ is isotopic to {some} branching map $m^\pm_i$.
\item The restrictions {$\d^\pm_1(\S')\colon [0,1] \to  \C(n)/S_n$} of $\S'$ to the left and right sides of $D^2$ {are each} constant paths.
\item {The restrictions $b_1=\d^+_2(\S')\colon [0,1] \to  \C(n)/S_n$ and $b_2=\d^-_2(\S')\colon [0,1] \to  \C(n)/S_n$  of $\S'$ to the top and bottom sides of $D^2$ each define braids.}
\end{enumerate}
\end{Definition}
\begin{Definition}
Two braided surfaces are equivalent if there exists a smooth homotopy between them which at each point is a braided surface.
\end{Definition}


\subsection{Arnold {Lemma} and Arnold basis}
{The basis  (defined by Arnold) of the cohomology ring of the configuration space $\C(n)$ described in this subsection will be crucial later, essentially leading to {Theorem} \ref{iffflat} below, and the definition of the differential crossed module of horizontal 2-chord diagrams in Section \ref{dcm}.}

In \cite{Ar}, Arnold addressed the cohomology ring (over the ring $\Z$ of integers) of the configuration space $\C(n)$ of $n$-particles in $\C$ (from which the results below can be inferred). Consider the following closed 1-forms:
$$\w_{ab}=\frac{dz_a-dz_b}{z_a-z_b}.$$
{These satisfy the following  relation (which is easy to prove), usually called Arnold's {Lemma}:}
\begin{equation}\label{AR}
\w_{ab}\wedge\w_{bc}+\w_{bc}\wedge\w_{ca}+\w_{ca}\wedge\w_{ab} = 0.
\end{equation}

Consider the graded commutative algebra of differential forms in $\C(n)$, with wedge product. Let $A_n$ be the subalgebra of it generated by the 1-forms $\w_{ab}$.  Then $A_n$ is isomorphic to the  (De Rham) cohomology ring of $\C(n)$, and  in particular a differential form in $A_n$ is zero if and only if it is cohomologous to zero. 

Basis for the degree 2 and 3 components of $A_n$ are, respectively: 
\begin{equation}\label{AB2}
\{ \w_{i_aj_a}\wedge\w_{i_bj_b} \, \mbox{ s.t. } \, i_k<j_k  \mbox{ and } j_k<j_{k^{\prime}} \mbox{ for } k<k^{\prime} \}.
\end{equation}
and
\begin{equation}\label{AB3}
\{ \w_{i_aj_a}\wedge\w_{i_bj_b} \wedge\w_{i_cj_c}\, \mbox{ s.t. } \, i_k<j_k \mbox{ and } j_k<j_{k^{\prime}} \mbox{ for } k<k^{\prime}\},
\end{equation}
where all indices run in $\{1,\ldots ,n\}$.

From this we can see that if $n=3$ and $n=4$ (respectively) then the following differential forms in $\C(n)$ are linearly independent (which can easily be proved directly):
$$\w_{12} \wedge \w_{13}, \quad \w_{12} \wedge \w_{23}$$
and 
 $$\w_{12} \wedge \w_{13}\wedge \w_{14}, \quad  \w_{12} \wedge \w_{23}\wedge \w_{14},  \quad  \w_{12} \wedge \w_{13}\wedge \w_{24}, \quad  \w_{12} \wedge \w_{23}\wedge \w_{24} \quad \w_{12} \wedge \w_{13}\wedge \w_{34}, \quad  \w_{12} \wedge \w_{23}\wedge \w_{34}.$$ 

\subsection{Flatness conditions for $\gl(\V)$-valued $2$-connections}\label{FC}

Let  $\V$ be a  chain complex of vector spaces. Recall the construction of the differential crossed module $\gl(\V)=\big ( \beta\colon \gl^1(\V) \to \gl^0(\V),\t \big )$ defined from $\V$, subsection \ref{dcmcc}.
 Consider a positive integer $n$. Suppose we have a representation $\s \in S_n \mapsto \rho_\s \in \Aut(\V)$ of the symmetric group $S_n$  on $\V$ by (necessarily invertible) chain maps. (The main example for this paper is the case when $\V$ is the tensor product of $n$ copies of a  chain complex $\Wc$, with the obvious action of $S_n$). Then $\rho_\s(f)=\rho_\s f \rho_\s^{-1}$ and $\rho_\s(s)=\rho_\s s \rho_s^{-1}$, for $f \in \gl^0(\V)$ and $s \in \gl^1(\V)$, define a representation of $S_n$ by differential crossed module maps $\gl(\V) \to \gl(\V)$. 

We are interested in flat local 2-connection pairs $(A,B)$ in $\C(n)$ with values in the differential crossed module $\gl(\V)$, such that the associated {two-dimensional} holonomy descends to a two-dimensional holonomy in $\C(n)/S_n$. Any map $\gamma \colon [0,1]\to  \C(n)/S_n$ can be lifted to $\C(n)$, and all liftings are related by the action of $S_n$ on $\C(n)$. So does any homotopy $\Gamma$ connecting paths $\gamma$ and $\gamma'$ in $\C(n)/S_n$. Therefore defining a 2-dimensional holonomy in $\C(n)/S_n$ directly from $(A,B)$ can be achieved if the  local 2-connection pair $(A,B)$ is invariant under the symmetric group $S_n$, in the sense that for each $\s \in S_n$
\begin{equation}\label{noquotient}
 \rho_\s^{-1}\big(\s^*(A)\big)=A  \textrm{ and } \rho_\s^{-1}\big(\s^*(B)\big)=B 
\end{equation}
where $\s\colon  \C(n) \to \C(n)$ denotes the obvious diffeomorphism given by $\s$.\footnote{There may be some space for relaxing the $S_n$ invariance of $(A,B)$ by considering non-trivial 2-vector bundles over $\C(n)/S_n$.}

In light of this discussion, let us start by addressing flat $\gl(\V)$-valued local 2-connection pairs in $\C(n)$. Consider a family of chain maps $\{r_{ab}\}\in\gl^0(\V)$ $(a,b \in \{1,\ldots ,n\}, \, a\neq b)$ such that
\begin{equation}
\label{r}
r_{ab} = r_{ba} , \qquad [r_{ab},r_{cd}] = 0 \; \mbox{ for } \{a,b\}\cap\{c,d\}=\emptyset
\end{equation}
and the closed differential forms
$$\w_{ab}=\frac{dz_a-dz_b}{z_a-z_b} $$
on $\C(n)$. We define a $\gl^0(\V)$-valued $1$-form  $A$ over $\C(n)$ as:
\begin{equation}
\label{Aconn}
A=\sum_{a<b}\w_{ab}r_{ab} .
\end{equation}

The curvature $\F_A=dA+\frac{1}{2} A\wedge^{{[-,-]}} A$  of $A$ is then (since the forms $\w_{ab}$ are closed):
$$ \F_A = \frac{1}{2} A\wedge^{{[-,-]}} A = \frac{1}{2} \sum_{i<j \, ; k<l} [r_{ij},r_{kl}] \, \w_{ij}\wedge\w_{kl} $$
and only the terms with one repeated index among $(i,j,k,l)$ contribute; by equation \eqref{r}. Considering the various cases we can write (these calculations appear in \cite{BN,Ka,Ko}):
\begin{equation*}
\begin{split}
\F_A & = \frac{1}{2} \big(\sum_{i<j<l} +\sum_{i<l<j}\big) [r_{ij},r_{il}] \, \w_{ij}\wedge\w_{il} + \frac{1}{2} \sum_{k<i<j} [r_{ij},r_{ki}] \, \w_{lj}\wedge\w_{kl} \,  \\ 
	 & \quad \quad + \frac{1}{2} \sum_{i<j<l} [r_{ij},r_{jl}] \, \w_{ij}\wedge\w_{jl} + \frac{1}{2} \big(\sum_{i<k<j} + \sum_{k<i<j} \big) [r_{ij},r_{kj}] \, \w_{ij}\wedge\w_{kj} \\
     & = \sum_{a<b<c} [r_{ab},r_{ac}] \, \w_{ab}\wedge\w_{ac} + [r_{ab},r_{bc}] \, \w_{ab}\wedge\w_{bc} + [r_{ac},r_{bc}] \, \w_{ac}\wedge\w_{bc} \, .
\end{split}
\end{equation*}
We express $\F_A$ along the Arnold basis of $2$-forms 
$$\{ \w_{i_aj_a}\wedge\w_{i_bj_b} \, \mbox{ s.t. } \, i_k<j_k \mbox{ and } j_k<j_{k^{\prime}} \mbox{ for } k<k^{\prime} \}$$ 
where all indices run in  $\{1,\ldots ,n\}$. By Arnold's lemma $\w_{ac}\wedge\w_{bc} = \w_{ab}\wedge\w_{bc} - \w_{ab}\wedge\w_{ac}$; therefore
$$ \F_A = \sum_{a<b<c} \big( \, [r_{ab},r_{ac}] - [r_{ac},r_{bc}] \, \big) \, \w_{ab}\wedge\w_{ac} + \big( \, [r_{ab},r_{bc}] + [r_{ac},r_{bc}] \, \big) \, \w_{ab}\wedge\w_{bc} $$      
and defining 
\begin{equation}
\label{VRdef}
\begin{split}
V_{abc} & = [r_{ab},r_{bc}] = r_{ab}r_{bc} - r_{bc}r_{ab} \\
R_{abc} & = V_{abc} - V_{bca} = [r_{ab}+r_{ac},r_{bc}]
\end{split}
\end{equation}
we eventually have
\begin{equation}
\label{FA}
\F_A = \sum_{a<b<c} R_{bac} \, \w_{ab}\wedge\w_{ac} + R_{abc} \, \w_{ab}\wedge\w_{bc} \, .
\end{equation}
Note that for the usual KZ-connection $V_{abc}=V_{bca}=V_{cab}$, which ensures flatness ($\F_A=0$).

We then need a $\gl^1(\V)$-valued 2-form $B$ such that $\beta(B)=\F_A$. We also want $(A,B)$ to be a flat 2-connection, so we impose the vanishing of the $2$-curvature $3$-form $\M_{(A,B)}$, see subsection \ref{2conn}.   
To match the the  condition $\beta(B)=\F_A$, we  define $B\in\Omega^2(\C(n),\gl^1(\V))$ as having the form:
\begin{equation}
\label{defB}
B = \sum_{a<b<c} K_{bac} \, \w_{ab}\wedge\w_{ac} + K_{abc} \, \w_{ab}\wedge\w_{bc} 
\end{equation}
for some $K_{abc}, K_{bac} \in\gl^1(\V)$ (where $1\leq a<b<c \leq n$), such that
\begin{equation}\label{kkkk} 
\b(K_{abc})=R_{abc} \textrm{ and } \b(K_{bac})=R_{bac}.
\end{equation}
We also suppose that:
\begin{equation}\label{lll} r_{ab} \act K_{ijk}=0 \textrm{ if } \{a,b\} \cap \{i,j,k\}=\emptyset.                                                                                                                                                                                                                                                                                                                    \end{equation}
 Given that $dB=0$, the curvature of $(A,B) $ is  $\M_{(A,B)}=A \wedge^\t B$. We compute the components of the $3$-form $A\wedge^{\triangleright}B$ along the Arnold basis
\begin{equation}
\label{3arn} 
\{ \w_{i_aj_a}\wedge\w_{i_bj_b} \wedge\w_{i_cj_c}\, \mbox{ s.t. } \, i_k<j_k  \mbox{ and } j_k<j_{k^{\prime}} \mbox{ for } k<k^{\prime}\}
\end{equation}
where all indices run in $\{1,\ldots ,n\}$. The vanishing of these components is equivalent to $2$-flatness $\M_{(A,B)}=0$.
By \eqref{lll}, only the terms where $\#\{i_a,j_a,i_b,j_b,i_c,j_c\}=4$ will affect the calculations.
\begin{Theorem}
\label{iffflat}
Given a $\gl(\V)$-valued $2$-connection $(A,B)$ on $\C(n)$ with $A$ as in \eqref{Aconn} and $B$ as in \eqref{defB}, the $2$-curvature $3$-form $\M_{(A,B)}$ vanishes, i.e. the 2-connection is flat, if and only if the following conditions are satisfied:
\begin{equation}\label{relfl}
\begin{split}
& r_{ad}\act (K_{bac} + K_{bcd}) + (r_{ab} + r_{bc} + r_{bd}) \act K_{cad} - (r_{ac}+r_{cd})\act K_{bad} = 0 \\
& r_{bd}\act (K_{abc} + K_{acd}) + (r_{ab} + r_{ad} + r_{ac}) \act K_{cbd} - (r_{bc}+r_{cd})\act K_{abd} = 0 \\
& r_{bc}\act (K_{bad} + K_{cad}) + r_{ad}\act (K_{cbd} + K_{bcd} -K_{abc}) = 0 \\
& r_{ac}\act (K_{abd} + K_{cbd}) + r_{bd}\act (K_{cad} + K_{acd} -K_{bac}) = 0 \\
& r_{cd}\act (K_{bac} + K_{bad}) + (r_{ab} + r_{bc} + r_{bd}) \act K_{acd} - (r_{ac}+r_{ad})\act K_{bcd} = 0 \\
& r_{cd}\act (K_{abc} + K_{abd}) + (r_{ab} + r_{ac} + r_{ad}) \act K_{bcd} - (r_{bd}+r_{bc})\act K_{acd} = 0 \\
\end{split}
\end{equation}
with $a<b<c<d \in \{1,\ldots ,n\}$. 
\end{Theorem}
Note that exchanging $a\leftrightarrow b$ in the first, third and fifth condition we get respectively the second, fourth and sixth. Exchanging $ a\leftrightarrow c$ in the first yields the fifth, if we also impose the condition $K_{bca}=K_{bac}$.\\

\begin{Remark}
 {As we will see in the proof of {Theorem} \ref{refnow}, the relations appearing in {Theorem} \ref{iffflat} are satisfied if we put $K_{abc}=[r_{ab}+r_{ac},r_{bc}]\in \gl^0(\V)$, where $\t$ is the adjoint action of $\gl^0(\V)$ on $\gl^0(\V)$. This turns out to be equivalent to Bianchi identity $d {\rm F}_A+A \wedge {\rm F}_A=0$, as read in the Arnold basis of the cohomology ring of the configuration space $\C(n)$, equation \eqref{AB3}.} 
\end{Remark}

\noindent
\begin{Proof} {\bf(Of {Theorem} \ref{iffflat})}
Writing explicitly $A\wedge^{\act}B$ (which we want to set to zero) we have 
$$ \sum_{i<j \, ; \, a<b<c } (r_{ij}\act K_{bac}) \, \w_{ij}\wedge\w_{ab}\wedge\w_{ac} + (r_{ij}\act K_{abc}) \, \w_{ij}\wedge\w_{ab}\wedge\w_{bc} \, .$$
The terms without repeated indices between $(i,j)$ and $(a,b,c)$ are zero because in that case the action of $r_{ij}$ on $K_{abc}$ vanishes, while the terms with $\{i,j\}\subset\{a,b,c\}$ are zero by antisymmetry of differential forms. Hence we only consider one repeated index: $i=a$, $i=b$, $i=c$ or the analogue three cases for $j$. Once we fix the repeated index, say $i=a$, we have a contribution along $\w_{aj}\wedge\w_{ab}\wedge\w_{ac}$ for the first term and along
$\w_{ai}\wedge\w_{ab}\wedge\w_{bc}$ for the second term. Next, we distinguish among the different relative orderings of $j$ with respect to $a,b,c$: we can have $a<j<b<c$, or $a<b<j<c$, or $a<b<c<j$. After we have made explicit all the possible cases for all the possible different repeated indices, we write everything along the Arnold basis of $3$-forms \eqref{3arn}. We have contributions only along elements with one repeated index, which correspond to the following linearly independent differential forms:
$$\w_{ab} \wedge \w_{ac} \wedge \w_{ad} ; \quad \w_{ab} \wedge \w_{bc} \wedge \w_{bd} ; \quad \w_{ab} \wedge \w_{bc} \wedge \w_{ad} ;\quad \w_{ab} \wedge \w_{ac}  \wedge \w_{bd} ;\quad  \w_{ab} \wedge \w_{ac} \wedge w_{cd} ; \quad \w_{ab}\wedge\w_{bc}\wedge\w_{cd}.$$
This leads therefore to six relations \eqref{relfl}, which appear in the same order as these  basis elements.
We compute in detail only the first relation, the others being similar.  (To simplify the notation in the rest of the proof we drop the wedge symbol among differential forms). Along $\w_{ab}\w_{ac}\w_{ad}$ we have contributions from: 
\begin{enumerate}[(i)]
\item $i=a$, first term and all possible intermediate positions of $j$: 
\begin{equation*}
\begin{split}
\sum_{a<j \, ; \, a<b<c} (r_{aj}\act K_{bac}) \, \w_{aj}\w_{ab}\w_{ac} & =  \big( \sum_{a<j<b<c} + \sum_{a<b<j<c} + \sum_{a<b<c<j} \big) (r_{aj}\act K_{bac}) \, \w_{aj}\w_{ab}\w_{ac} \\  
& = \sum_{a<b<c<d} ( r_{ab}\act K_{cad} - r_{ac}\act K_{bad} + r_{ad}\act K_{bac} ) \, \w_{ab}\w_{ac}\w_{ad}
\end{split}
\end{equation*} 
\item $j=b$, first term and $a<i<b$: 
\begin{equation*}
\sum_{a<i<b<c} (r_{ib}\act K_{bac}) \, \w_{ai}\w_{ab}\w_{ac} = \sum_{a<b<c<d} (r_{bc}\act K_{cad}) \, \w_{ab}\w_{ac}\w_{ad}
\end{equation*}
\item $j=c$, first term and $a<i$:
\begin{equation*}
\begin{split}
 \sum_{a<i \, ; \, a<b<c} (r_{ic}\act K_{bac}) \, \w_{ai}\w_{ab}\w_{ac}  & =  \big( \sum_{a<i<b<c} + \sum_{a<b<i<c} \big) (r_{ic}\act K_{bac}) \, \w_{ai}\w_{ab}\w_{ac} \\ 
 & = \sum_{a<b<c<d} ( r_{bd}\act K_{cad} - r_{cd}\act K_{bad} ) \, \w_{ab}\w_{ac}\w_{ad}
\end{split}
\end{equation*}
\item $j=c$, second term and $i<a$:
\begin{equation*}
\sum_{i<a<b<c} (r_{ic}\act K_{abc})\, \w_{ia}\w_{ib}\w_{ic} = \sum_{a<b<c<d} (r_{ad}\act K_{bcd})\, \w_{ab}\w_{ac}\w_{ad}
\end{equation*}
\end{enumerate} 
The sum of these contributions along $\w_{ab}\w_{ac}\w_{ad}$ is the first relation in \eqref{relfl}. 
\end{Proof} \\

In light of the discussion in the beginning of this subsection, let now us impose relations \eqref{noquotient}. These imply that for any permutation $\s \in S_n$ we must have:
$$\rho_\s\big({r_{ab}}\big)=r_{\s(a)\s(b)} .$$
Let $\tau_{ab}\in S_n$ be the transposition that exchanges $a$ and $b$. By imposing that $\tau_{ab}^*(B)=\rho_{\tau_{ab}}(B)$, we obtain the following conditions, by direct  calculations in the Arnold basis \eqref{AB3}:
\begin{align}
\rho_{\tau_{ab}}(K_{abc})&=K_{bac}  & \rho_{\tau_{bc}}(K_{bac})&=-K_{bac}-K_{abc} &\rho_{\tau_{bc}}(K_{abc})&=K_{abc} \nonumber\\\label{cond}
& & \rho_{\tau_{ac}}(K_{abc})&=-K_{abc}-K_{bac} & \rho_{\tau_{ac}}(K_{bac})&=K_{bac} 
  \end{align}
Indeed, fix $a<b<c$. For the case of the transposition $\tau_{bc}$, note that (we use Arnold's Lemma \eqref{AR}):
\begin{align*}
\tau_{bc}^*\left( K_{bac} \, \w_{ab}\wedge\w_{ac} + K_{abc} \, \w_{ab}\wedge\w_{bc} \right)&= K_{bac} \, \w_{ac}\wedge\w_{ab} + K_{abc} \, \w_{ac}\wedge\w_{cb}\\
&= -(K_{bac}+K_{abc})  \, \w_{ab}\wedge\w_{ac} +K_{abc} \, \w_{ab}\wedge\w_{bc}.
\end{align*}
This is just the projection of $\tau_{bc}^*(B)$ along the basis elements $\w_{ab}\wedge\w_{ac}$ and $\w_{ab}\wedge\w_{bc}$. The projection of $\rho_{\tau_{bc}}(B)$ along these is:
$$\rho_{\tau_{bc}}(K_{bac}) \, \w_{ab}\wedge\w_{ac} + \rho_{\tau_{bc}}(K_{abc}) \, \w_{ab}\wedge\w_{bc}.$$

Conditions \eqref{cond} permit us to say what $K_{ijk}$ should be when we do not have $i,j<k$. If $a<b<c$ we put, in function of the given $K_{abc}$ and $K_{bac}$:
\begin{align}
K_{cab}&=-K_{bac}-K_{abc} & K_{acb}&=K_{abc}\label{AAaa}\\
  K_{cba}&=-K_{abc}-K_{bac} & K_{bca}&=K_{bac}\label{BB} 
\end{align}
Note that for each distinct $i,j,k$ we have:
$$K_{ijk}+K_{jki}+K_{kij}=0 \textrm{ and also } R_{ijk}=R_{ikj}. $$
Also for each distinct $i,j,k$ and permutation $\s$ of $\{i,j,k\}$ we have $$K_{\s(i)\s(j) \s(k)}=\rho_{\s} (K_{ijk}).$$

By looking at the coefficients, in the Arnold basis, of both sides of the equation $\rho_{\sigma}^{-1}\big(\s^*(B)\big)=B$  it is easy to see, given any transposition $\s$ of $\{1,\ldots ,n\}$, that   the condition $\rho_{\sigma}^{-1}\big(\s^*(B)\big)=B$ implies that we must have $K_{\s(i)\s(j)\s(k)}=\rho_{\s} (K_{ijk})$.

Now note:
\begin{align}
B& = \sum_{a<b<c} K_{cba} \, \w_{bc}\wedge\w_{ba} + K_{bca} \, \w_{bc}\wedge\w_{ca}\\
 & = \sum_{a<b<c} K_{acb} \, \w_{ca}\wedge\w_{cb} + K_{cab} \, \w_{ca}\wedge\w_{ab}.
\end{align}
By considering these expressions of $B$ together with \eqref{defB}, putting $\Omega_{abc}=\w_{ab} \wedge \w_{bc}$ we have that:
\begin{equation}\label{Simp}
B= \frac{1}{3}\sum_{a,b,c} K_{abc} \, \Omega_{abc}. 
\end{equation}

From \eqref{Simp} it is clear that if we have $\rho_{\s}\big({K_{abc}}\big)=K_{\s(a)\s(b)\s(c)} $, now for any permutation $\s\in S_n$, then it follows the desired invariance $\rho_\s^{-1}\big(\s^*(B)\big)=B$. 

We have proven:
\begin{Lemma}
Let $\V$ be a chain complex. Consider a representation $\s \mapsto \rho_\s$ of $S_n$ on $\V$ by chain-complex isomorphisms. Choose chain complex maps $r_{ab}\in \gl_0(\V)$, where $a,b \in \{1,\dots,n\}$, with $r_{ab}=r_{ba}$ and $a\neq b$, and also chain-homotopies (up to 2-fold homotopy) $K_{ijk}\in \gl^1(\V)$, where $i,j,k$ are distinct indices in $\{1,\dots,n\}$. There are to satisfy \eqref{r}, \eqref{kkkk} and \eqref{lll}. The $\gl(\V)$-valued 2-connection $(A,B)$, where 
$$A=\sum_{a<b}\w_{ab}r_{ab} \quad\textrm{and}\quad  B = \sum_{a<b<c} K_{bac} \, \w_{ab}\wedge\w_{ac} + K_{abc} \, \w_{ab}\wedge\w_{bc} $$
has zero {curvature 3-form}, being, further, invariant under the action of the symmetric group $S_n$ if and only if conditions \eqref{relfl} are satisfied (with $a<b<c<d$) and, moreover, for each distinct $i,j,k$ we have:
\begin{equation}
K_{ijk}+K_{jki}+K_{kij}=0, \quad  K_{ijk}=K_{ikj},
\end{equation}
and  for each permutation $\s$ of $\{1,\dots,n\}$ we have 
\begin{equation}\label{asn}
r_{\sigma(i) \sigma(j)}=  \rho_{\sigma}( r_{ij})  \textrm{ and also }  K_{\s(i)\s(j) \s(k)}=\rho_{\s} (K_{ijk}).
\end{equation}
\end{Lemma}

Now note that given that the maps $\rho_\sigma\colon \V \to \V$ are chain complex maps, they  induce morphisms of crossed modules $\gl(\V) \to \gl(\V)$.  Therefore, if we suppose that equation \eqref{asn} holds, then if one of the equations of \eqref{relfl} is true then so is any equation obtained from it by permuting indices. By using the comments just after {Theorem} \ref{iffflat} it follows:
\begin{Theorem}
\label{thm13}
 In the conditions of the previous lemma $(A,B)$ is flat and invariant under the action of $S_n$ if and only if for any  $a<b<c<d \in \{1,\dots,n\}$ we have
\begin{equation}\label{mnb}
\begin{split}
& r_{ad}\act (K_{bac} + K_{bcd}) + (r_{ab} + r_{bc} + r_{bd}) \act K_{cad} - (r_{ac}+r_{cd})\act K_{bad} = 0 \\
& r_{bc}\act (K_{bad} + K_{cad}) -r_{ad}\act (K_{dbc} +K_{abc}) = 0,
\end{split}
\end{equation}
also
\begin{equation}
K_{abc}+K_{bca}+K_{cab}=0, \quad \quad \quad K_{bca}=K_{bac} \, ,
\end{equation}
and for each permutation $\sigma \in S_n$:
\begin{align}r_{\sigma(a) \sigma(b)}=  \rho_{\sigma}( r_{ab})\textrm{ and } K_{\s(a)\s(b) \s(c)}=\rho_{\s} (K_{abc}).
 \end{align}
Moreover in this case equations \eqref{mnb} hold for any permutation of the indices. 
\end{Theorem}

The interpretation of these conditions (for $2$-flatness and $S_n$-invariance) in {terms} of (a categorified version of) chord diagrams will be the subject of the following sections.


\subsection{Free differential crossed modules}

Consider a Lie algebra $\g$, a set $S$ and a map $\d_0\colon S\rightarrow \g$. The aim of this section is to define the free differential (pre)crossed module over this data. As usual it will be defined by a universal property, and a model for it will be presented. For details on the construction of free crossed modules of groups see \cite{BHS,B}. For the incorporation of additional relations see \cite{FM1}.

The first step is the notion of free $\g$-Lie algebra, where a $\g$-Lie algebra is a Lie algebra with a $\g$-action by derivations. 
Recall that, given a vector space $V$, the free Lie algebra $F(V)$ over $V$ is a Lie algebra $F(V)$, together with a linear inclusion map $i\colon V \to F(V)$, such that for any Lie algebra $L$, any linear map $g\colon V \to L$ extends uniquely to a lie algebra map $g'\colon F(V) \to L$. The Lie algebra $F(V)$ can be constructed from the tensor algebra  $T(V)$ of $V$ (with the usual commutator of an associative algebra) by considering the Lie subalgebra $F(V)$ of it generated  by $V$. Note that  any linear map $f\colon V \to V$ extends uniquely to an algebra derivation of the tensor algebra $T(V)$, and therefore to a Lie algebra derivation of $F(V)$.

\begin{Definition}
Let $\g$ be a Lie algebra, $S$ a set. The free $\g$-Lie algebra over $S$ is a $\g$-Lie algebra $F_{\g}(S)$ together with a set map $i:S\rightarrow F_{\g}(S)$ with the following universal property: for any $\g$-Lie algebra $M$ and any map $f:S\rightarrow M$ there exists a unique $\g$-Lie algebra morphism $f^{\prime}:F_{\g}(S)\rightarrow M$ such that $f^{\prime}i = f$. 
\end{Definition}

To exhibit a model for $F_{\g}(S)$ we use the universal enveloping algebra $\U(\g)$, remember that on every $\g$-module is induced a unique $\U(\g)$-module structure. Consider the free Lie algebra $F_{\g}(S)$ on the vector space $\U(\g)\cdot S := \oplus_{s \in S} \U(\g)$. Denote the  elements of  $\U(\g)\cdot S $ as $(u,a)$, $u\in \U(\g)$ and $a\in S$, and define $i(a)=(1,a)\in \U(\g)\cdot S\subset  F_{\g}(S)$, where $a \in S$. Define the $\g$-action as $X\act(u,a):=(Xu,a)$, and extend it as a Lie algebra derivation. 

\begin{Proposition}
Given any $\g$-Lie algebra $M$ together with a map $f:S\rightarrow M$ there exists a unique $\g$-Lie algebra morphism $f^{\prime}: F_{\g}(S) \rightarrow M$ such that $f^{\prime}i = f$.
\end{Proposition}
\begin{Proof}
Since (by PBW {Theorem}) $\U(\g)\cdot S$ is generated by the $\g$ action on $i(S)$, a map with such properties is clearly unique. For existence, consider the Lie algebra map $f'\colon F_{\g}(S) \to M$, given by the linear map $f''\colon \U(\g)\cdot S \to M$ such that 
$f''(u,a)=u \act f(a)$. 
This completes the proof.
\end{Proof}

From the previous definition and proposition, it is clear that looking at $\g$ itself as a $\g$-Lie algebra (with $\g$ action given by Lie bracket), for every map $\d_0:S\rightarrow\g$ we have a differential pre-crossed module $\d:F_{\g}(S)\rightarrow \g$. It satisfies the following universal property, and for this reason it is referred to as the free differential pre-crossed module over set map $\d_0:S\rightarrow\g$.

\begin{Proposition}
\label{freedpcm}
Given a Lie algebra $\g$, a set $S$ and a set map $\d_0:S\rightarrow \g$, the differential pre-crossed module $\d:F_{\g}(S)\rightarrow \g$ has the following universal property: for any differential pre-crossed module $\d':\h\rightarrow\g$ and any map $t:S\rightarrow\h$ such that $\d_0=\d' \, t$ there exists a unique $\g$-Lie algebra morphism $\alpha:F_{\g}(S)\rightarrow\h$, extending $t$, and  such that $\d=\d' \, \alpha$.
\end{Proposition}
\begin{Proof} 
Uniqueness is trivial. For existence, consider the unique $\g$-Lie algebra map $\alpha\colon F_\g(S) \to \h$ extending $t$. It is trivial that $\d=\d' \, \alpha$ since this is true for the set $S$ generating $F_\g(S)$ as a $\g$-Lie algebra.
\end{Proof} \\

We project to differential crossed modules by adding the (differential) Peiffer relation \eqref{sid}. Given a differential pre-crossed module $\d:\h\rightarrow\g$ this amounts to quotient $\h$ by the Peiffer ideal $\Pf\subset \h$ generated by elements of the form $\pf(\xi,\nu)= \d(\xi)\act\nu - [\xi,\nu]$ for all possible  $\xi,\nu \in \h$.

\begin{Proposition}
\label{pfquot}
Given a differential pre-crossed module $\d:\h\rightarrow\g$ and denoting $\h^{\Pf}:= \h/\Pf$, the induced $\g$-action and $\d$ map on the quotient make $\d:\h^{\Pf}\rightarrow\g$ a differential crossed module.
\end{Proposition}
\begin{Proof}
We need to prove is that $\Pf$ is stable for the $\g$-action, i.e. $\g\act\Pf\subset\Pf$, and that $\d(\Pf)=0$. It is sufficient to check both properties on generators; for any $X\in\g$ and $ \xi,\mu \in \h$ we have
\begin{equation*}
\begin{split}
X\act \pf(\xi,\nu) & = X\act \big((\d\xi)\act\nu - [\xi,\nu] \big) = [X,\d(\xi)]\act\nu + \d(\xi) \act (X\act\nu) - [X\act\xi,\nu] - [\xi,X\act\nu] \\
 & = (\d(X\act\xi))\act\nu - [X\act\xi,\nu] + \d(\xi)\act(X\act\nu) - [\xi,X\act\nu] = \pf(X\act\xi,\nu) + \pf(\xi,X\act\nu) \\
\d(\pf(\xi, \nu)) & = \d \big((\d\xi)\act\nu - [\xi,\nu] \big) = [\d(\xi),\d(\nu)] - [\d(\xi),\d(\nu)] = 0 \, .
\end{split}
\end{equation*}
This completes the proof.
\end{Proof} \\    

It is natural to adapt the notion of free differential pre-crossed module to the differential crossed module case.

\begin{Definition}[Free differential crossed module]
\label{fdxmod}
Given a Lie algebra $\g$, a set $S$ and a map $\d_0:S\rightarrow\g$ the free differential crossed module over $(S,\d_0)$ is a differential crossed module $\d:\fxmod\rightarrow\g$ together with a set map $i:S\rightarrow \fxmod$ such that $\d_0= \d i$, satisfying the following universal property: for every differential crossed module $\d':\h\rightarrow\g$ and map $t:S\rightarrow\h$ such that $\d_0=\d' \, t$ there exists a unique morphism $\alpha:\fxmod\rightarrow\h$ of $\g$-Lie algebras, extending $t$, such that $\d = \partial'\alpha$.
\end{Definition}

A model for $\fxmod$ can be obtained from the free differential pre-crossed module $\d:F_{\g}(S)\rightarrow\g$ by considering the quotient $F_{\g}(S)/\Pf$. By the results of Propositions \ref{freedpcm} and \ref{pfquot} it is easy to verify that $\d: F_{\g}(S)/\Pf \rightarrow\g$ satisfies the universal property of Definition \ref{fdxmod}.


\subsection{The differential crossed module of 2-chord diagrams}\label{dcm}

We start from the usual Lie algebra of horizontal chord diagrams $\ch_n$ considered in the introduction (see also below), and remove the 4-term  relations \eqref{j1}, obtaining a larger algebra $\ch^+_n$. The Lie algebra $\ftch_n$ generated by the 4-term relations (divided by the crossed module relations) is lifted to appear in  a differential crossed module $\d:\ftch_n\rightarrow\ch^+_n$. We then consider the quotient of $\ftch_n$ by a set of higher order relations (implying 2-flatness) obtaining a new differential crossed module 
\begin{equation}
\label{2chd} {2\mathfrak{ch}_n}=(
\d:\tch_n\rightarrow\ch^+_n )\, .
\end{equation}
The geometrical interpretation of   ${2\mathfrak{ch}_n}$, coming from the discussion in subsection \ref{FC}, justifies the  name \textit{differential crossed module of totally symmetric horizontal $2$-chord diagrams} for \eqref{2chd}.

\begin{Definition}[Algebra of horizontal chord {diagrams}]
Fix $n\in\mathbb{N}$. The {Lie} algebra of horizontal chord {diagrams} $\ch_n=L(r_{ab})/J$ is the  Lie algebra freely generated by the symbols $r_{ab}$, $a\neq b, \, a,b\in\{1,\ldots ,n\}$, modulo the ideal $J$ generated by the following relations:
\begin{align}
\label{j0}
& r_{ab} = r_{ba} ,  & [r_{ab},r_{cd}] = 0 \; \mbox{ for } \{a,b\}\cap\{c,d\}=\emptyset \, , \\
\label{j1}
& [r_{ab}+r_{ac},r_{bc}]  = R_{abc} = 0  \, .&  
\end{align}
\end{Definition}
The relation (\ref{j1}) will be called the 4-term relation \cite{BN,Ka,Ko}.

The differential form 
\begin{equation}\label{defA} A=\sum_{1\leq a <b \leq n} \w_{ab} r_{ab}\end{equation}  defines a flat connection in the trivial vector bundle $\C(n) \times \ch_n$, in other words $d A+\frac{1}{2}A\wedge A=0$. This is well known and follows from the calculation in the beginning of subsection \ref{FC}. Consider the action of the symmetric group $S_n$ on the Lie algebra $\ch_n$ defined on generators as  $\rho_\s({r_{ab}})=r_{\s(a) \s(b)}$ (clearly this is a Lie algebra morphism.) Consider the product action of $S_n$ on  $\C(n) \times \ch_n $. Then $A$ is invariant under this action, and therefore defines a connection (also denoted with $A$) on the vector bundle $\big( \C(n) \times \ch_n\big)/S_n$, over $\C(n)/S_n$. 

Given a positive integer $n$, we now want to find a differential crossed module $2 \mathfrak{ch}_ n$, the differential 2-crossed module of (totally symmetric) horizontal {2-chord} diagrams, together with a flat local 2-connection  pair $(A,B)$ in $\C(n)$ with values in $2\mathfrak{ch}_n .$
To deal with $2$-connections and $2$-flatness, we are  interested in weakening condition \eqref{j1}. We denote $J_0$ the ideal generated by \eqref{j0} alone, and consider the larger algebra $\ch^+_n:=L(r_{ab})/J_0$. In particular, we use the `removed' relations $R_{abc}=0$ to construct a differential crossed module over $\ch^+_n$. Note that $r_{ab}=r_{ba}$ implies $R_{abc}=R_{acb}$.

\begin{Definition}
Fix  $n\in\mathbb{N}$. Let $K$ be the set
$$K= \{K_{abc}, \, a,b,c \in \{1,\ldots ,n\}, \, a\neq b, \, a\neq c, \, b\neq c \}$$
and consider the map $\d_0:K\rightarrow\ch_n^+$ sending $K_{abc}$ into $R_{abc}$. The differential crossed module of free horizontal $2$-chord diagrams is the free differential crossed module over $(K,\d_0)$. It will be denoted as $\d:\ftch_n\rightarrow\ch^+_n$.
\end{Definition}

The geometrical meaning of this construction is that when the connection $1$-form \eqref{defA} takes values in $\ch^+_n$ instead of $\ch_n$, it is no longer flat. We can however recover flatness at the level of a $2$-connection, proceeding as follows.

\begin{Theorem}[(Totally symmetric) horizontal 2-chord diagrams]  \label{refnow}
Define $J_2\subset\ftch_n$ to  be the $\ch^+_n$-module generated by the relations:
\begin{equation}
\label{j2}
\begin{split}
& r_{ad}\act (K_{bac} + K_{bcd}) + (r_{ab} + r_{bc} + r_{bd}) \act K_{cad} - (r_{ac}+r_{cd})\act K_{bad} = 0 \\
& r_{bc}\act (K_{bad} + K_{cad}) -r_{ad}\act (K_{dbc} +K_{abc}) = 0,
\end{split}
\end{equation}
also
\begin{equation}
K_{abc}+K_{bca}+K_{cab}=0 \quad \quad \quad K_{bca}=K_{bac} \, ,
\end{equation}
and of course
\begin{align}\label{commun}
 r_{ab} \t K_{a'b'c'}&=0 \quad \textrm{ if } \{a,b\} \cap \{a',b',c'\}= \emptyset.
\end{align}
Then $\d(J_2)=0$, so that $J_2$ is an ideal in $\ftch_n$, $\d$ is well defined on the quotient $\tch_n = \ftch_n/J_2$ and $\d:\tch_n\rightarrow\ch^+_n$ is a differential crossed module, referred to as the differential crossed module of (totally symmetric) horizontal 2-chord diagrams $\tchmf_n$.
\end{Theorem}
Note that $\rho_\s(K_{abc})=K_{\s(a)\s(b)\s(c)}$ and $\rho_\s(r_{ab})=K_{\s(a)\s(b)}$ defines an action of $S_n$ on  $\tchmf_n$ by differential crossed module maps. From relation \eqref{commun} and the definition of a differential crossed module it also follows that:
\begin{equation}
 [K_{abc}, K_{a'b'c'}]=0 \quad \textrm{ if } \{a,b,c\} \cap \{a',b',c'\}= \emptyset.
\end{equation}

\noindent \begin{Proof}
It is enough to compute $\d$ on the generators of $J_2$. We start with the first relation:
\begin{equation*}
\begin{split}
[r_{ad}, & R_{bac}+R_{bcd}] + [r_{ab}+r_{bc}+r_{bd},R_{cad}]- [r_{ac}+r_{cd},R_{bad}]  \\
 = & \, [r_{ad},[r_{ab}+r_{bc},r_{ac}] +[r_{bc}+r_{bd},r_{cd}]] + [r_{ab}+r_{bc}+r_{bd},[r_{ac}+r_{cd},r_{ad}]] - [r_{ac}+r_{cd},[r_{bd}+r_{ab},r_{ad}]]  \\
 = & \, [r_{ad},[r_{ab},r_{ac}]] + [r_{ad},[r_{bc},r_{ac}]] + [r_{ad},[r_{bc},r_{cd}]] + [r_{ad},[r_{bd},r_{cd}]] + [r_{ab},[r_{ac},r_{ad}]] + [r_{ab},[r_{cd},r_{ad}]] \,  \\
 & \, + [r_{bc},[r_{ac},r_{ad}]] + [r_{bc},[r_{cd},r_{ad}]] + [r_{bd},[r_{ac},r_{ad}]] + [r_{bd},[r_{cd},r_{ad}]] - [r_{ac},[r_{bd},r_{ad}]] - [r_{ac},[r_{ab},r_{ad}]] \,  \\
 & \, - [r_{cd},[r_{bd},r_{ad}]] - [r_{cd},[r_{ab},r_{ad}]] \, .
\end{split}
\end{equation*} 
Now we look separately at the terms which contain the same three pairs of indices:
\begin{itemize}
\item[-] indices $(ab)(ac)(ad)$: zero by Jacobi
\item[-] indices $(ac)(ad)(bc)$: by Jacobi the sum is $[[r_{ad},r_{bc}],r_{ac}]=0$
\item[-] indices $(ad)(bc)(cd)$: by Jacobi the sum is $[[r_{ad},r_{bc}],r_{cd}]=0$
\item[-] indices $(ad)(bd)(cd)$: zero by Jacobi
\item[-] indices $(ab)(ad)(cd)$: by Jacobi the sum is $[[r_{ab},r_{cd}],r_{ad}]=0$
\item[-] indices $(ac)(ad)(bd)$: by Jacobi the sum is $[[r_{bd},r_{ac}],r_{ad}]=0$
\end{itemize}
For the second relation the computation is similar: once the $R$ terms are made explicit, we simplify by using Jacobi identity. The remaining relations follow immediately.

By the definition of a differential crossed module, this implies that $J_2$ is in the center of $\ftch$, hence an ideal:
$$ [j,x]= \d(j)\t x = 0 \quad \forall j\in J_2 , \, x\in\ftch \, .$$
The rest of the statement now easily follows.  
\end{Proof}  \\

{By construction and the calculations in subsection \ref{FC}, we have the following theorem, which is the main result of this paper:}
\begin{Theorem}
 
The pair of forms with values in $ {2\mathfrak{ch}_n}=(\d:\tch_n\rightarrow\ch^+_n )$
\begin{equation}\label{defineAB}
A=\sum_{a<b} r_{ab}\, \w_{ab}  , \qquad B = \sum_{a<b<c} K_{bac} \, \w_{ab}\wedge\w_{ac} + K_{abc} \, \w_{ab}\wedge\w_{bc}
\end{equation}
defines a flat 2-connection pair in $\C(n)$, i.e. $\d(B)=\F_A$ and $d B+A \wedge^\t B=0$. Moreover, $(A,B)$ is invariant under the natural action of $S_n$.
\end{Theorem}
{The 2-connection $(A,B)$ defined in the previous theorem is our proposal for a categorified version of the Knizhnik-Zamolodchikov connection.}

\begin{Corollary}
\label{2chfl}
Let $\G=(\d\colon H \to G,\t)$ be a Lie crossed module, with associated differential crossed module $\mathfrak{G}=(\d\colon \le \to \g,\t)$. Suppose that $\mathfrak{G}$ is provided with an action of the symmetric group $S_n$ by differential crossed module maps.
For any morphism of crossed modules $\rho:\tchmf_n\rightarrow\mathfrak{G}$, preserving the action of the symmetric group, the $\mathfrak{G}$-valued $2$-connection $(A,B)$ over the configuration space of $n$ points $\mathbb{C}(n)$  defined as
$$ A=\sum_{a<b}\rho(r_{ab})\, \w_{ab}  , \qquad B = \sum_{a<b<c} \rho(K_{bac}) \, \w_{ab}\wedge\w_{ac} + \rho(K_{abc}) \, \w_{ab}\wedge\w_{bc} $$
where $\w_{ab}=\frac{dz_a-dz_b}{z_a-z_b}$, is flat. Moreover,  its {two-dimensional} holonomy descends to a {two-dimensional} holonomy over $\C(n)/S_n$, taking values in  $\G$.\end{Corollary}


\section{{Categorical representations of differential crossed modules and infinitesimal 2-R-matrices}}

{In this section, we present a Lie algebra framework in which flat 2-connections constructed in the realm of Corollary  \ref{2chfl} naturally fit. We intend to define the concept of an infinitesimal 2-R-matrix (categorifying the notion of an infinitesimal R-matrix $r \in \g'\tn \g'$, for  a Lie algebra $\g'$, see the Introduction), in a differential crossed module $\mathfrak{G'}=(\d\colon \e' \to \g',\t)$,   to be a pair of tensors $P$ and $r$, living in a quotient of the  tensor algebra of the underlying chain complex of $\mathfrak{G'}$, which satisfy analogous relations to the ones of {Theorem} \ref{refnow}.}

Given a chain complex $\V$ of vector spaces, the Lie  crossed module $\G$ appearing in Corollary  \ref{2chfl} will be of the form $\GL(\V^{ \btn n })=\big ( \beta\colon \GL^1(\V^{ \btn n }) \to \GL^0(\V^{\btn n}),\t \big )$, see Section \ref{lcomp}, where  $\V^{ \btn n }$ denotes the tensor product of $\V$ with itself $n$ times.

{Passing from elements in the differential crossed module $\mathfrak{G'}$ to elements in the differential crossed module $\gl(\V^{\btn n})$ makes heavy use of the notion of a representation of a differential crossed module in a chain complex of vector spaces, and the fact that these representations can be tensored. }

\subsection{Chain complexes and categorical representations of differential crossed modules}
\label{seccatreps}

Recall the construction of the differential crossed module $\gl(\V)=\big ( \beta\colon \gl^1(\V) \to \gl^0(\V),\t \big )$ defined from a chain complex $\V$ of vector spaces, subsection \ref{dcmcc}.

\begin{Definition}
Let $\mathfrak{G}$ be a differential crossed module. Let also $\V$ be a complex of vector spaces. A categorical representation $\rho$ of $\mathfrak{G}$ on $\V$ is a crossed module morphism $\rho=(\rho_1,\rho_0) \colon \mathfrak{G} \to \gl(\V)$.
\end{Definition}
For the case of length two chain complexes, this appeared for example in \cite{Po,FB}. The  following natural example appears in \cite{Wo}.

\begin{Example}[Adjoint representation]
Let $(\d\colon \e \to \g,\t)$ be a differential crossed module. The adjoint representation of $\mathfrak{G}$ on its underlying chain complex  is given by the pair $\rho=(\rho_1,\rho_2)$, where: 
\begin{itemize}
 \item If $X\in \g$ the chain map $\rho_0^X\colon \mathfrak{G} \to \mathfrak{G}$ is such that $$\rho_0^X(Y)=[X,Y]$$ and $$\rho_0^X(\zeta)=X \t \zeta$$ where $Y \in \g$ and $\zeta \in \e$.
\item If $\zeta \in \e$ the homotopy $\rho_1^\zeta\colon \g \to \e$ is such that 
$$\rho_1^\zeta(X)=-X \t \zeta. $$  
\end{itemize}
\end{Example}

It is an instructive exercise to prove this; clearly $\rho_0^{[X,Y]}=[\rho_0^X,\rho_0^Y]$, and by the crossed module rules $$\rho^{\d(\xi)}_0=\beta\big (\rho^\xi_1\big).$$
Also
\begin{align*}
 [\rho_1^\xi,\rho_1^\zeta](X)&=\d(X \t \zeta) \t \xi- \d(X \t \xi) \t \zeta =[X,\d(\zeta)] \t \xi-[X,\d(\xi)] \t \zeta  \\
 &=X \t \big (\d(\zeta) \t \xi\big)-\d(\zeta) \t\big( X \t \xi\big)-X \t \big (\d(\xi) \t \zeta\big)+\d(\xi) \t\big( X \t \zeta\big)\\
 &=X \t [\zeta,\xi]-[\zeta, X \t \xi]-X \t [\xi,\zeta]+[\xi,X \t \zeta] = X \t [\zeta,\xi]=\rho_1^{[\xi,\zeta]}(X)
\end{align*}
where the penultimate equation follows since $\g$ acts on $\e$ by derivations, which makes the last three terms cancel out.


\subsection{Tensoring categorical representations}\label{sectn}

\subsubsection{Tensor product of chain complexes}

For details on the tensor product of chain complexes see \cite{D}. Recall again  the construction of the differential crossed module $\gl(\V)=\big ( \beta\colon \gl^1(\V) \to \gl^0(\V),\t \big )$ defined from a chain complex $\V$ of vector spaces, subsection \ref{dcmcc}.
 Given chain complexes $\V=(V_i,\d)$ and $\Wc=(W_i,\d)$, the degree $n$ part of the tensor product $\U=\V\btn \Wc$ is:
$$U_n=\bigoplus_{i+j=n} V_i \tn W_j.$$
Given $x_i \in V_i$ and $y_j \in W_j$ we put
$$\d (x_i \tn y_j)=\d(x_i) \tn y_j+(-1)^i x_i \tn \d(y_j).$$
The complexes $\V\btn \Wc$ and $\Wc\btn \V$ are isomorphic, the isomorphism having the form $x_i \tn y_j \mapsto (-1)^{ij} y_j \tn x_i$.

Given $f \in \Hom^m(\V)$ and $g\in \Hom^n(\Wc)$ then $f \btn g \in \Hom^{m+n}(\V \btn \Wc)$, which has degree $m+n$, is defined as (for $x_i \in V_i$ and $y_j\in W_j$)
$$(f \btn g)(x_i \tn y_j)=(-1)^{ni} f(x_ i) \tn g(y_ j). $$
Therefore, if $f'$ and $g'$ have degrees $m'$ and $n'$ we have: 
$$(f \btn g) (f' \btn g')=(-1)^{m'n} (ff'\btn gg').$$

\begin{Lemma}
\label{A}
{If $f\colon \V \to \V$ and $g \colon \Wc \to \Wc$ are chain maps (of degree $0$) and $s\in \Hom^1(\V)$, $t \in \Hom^1(\Wc)$ are homotopies we have:}
$$\b(f \btn t)=f \btn \b(t) $$
and 
$$\b(s \btn g)=\b(s) \btn g \, .$$ 
\end{Lemma}
This result fails to hold if $f$ or $g$ are solely degree-$0$ maps (without being, further, chain maps).

\begin{Lemma} {If $s\in \Hom^1(\V)$ and $t \in \Hom^1(\Wc)$ are homotopies:}
 $$\b'(s \btn t)=s \btn \b(t)-\b(s) \btn t  .$$
\end{Lemma}
(Note $\b \b' (s\btn t)=0$, as it should.)

\begin{Corollary}\label{simp} {If $s\in \Hom^1(\V)$ and $t \in \Hom^1(\Wc)$ are homotopies, then as
 elements} of $${\gl^1(\V \btn \Wc)\doteq \Hom^1(\V \btn \Wc)/\b'(\Hom^2(\V \btn \Wc) )}$$ {the homotopies $\b(s) \btn t$ and $s \btn \b(t)$ coincide (in other words  they are the same up to 2-fold homotopy). }
\end{Corollary}

\begin{Lemma}
Let $k\in \Hom^2(\V)$ and $h \in \Hom^2({\cal W})$. Let $f\colon \V \to \V$ and $g\colon {\cal W} \to {\cal W}$ be chain maps. We have:
$$\b'(k \btn g)=\b'(k) \btn g \, , $$
$$\b'(f \btn h)=f \btn \b'(h). $$
\end{Lemma}
We therefore have, for example, {if $s,t \in \Hom^1(\V)$}:
$$\b'( ts \btn 1)=\b'(ts) \btn 1  .$$

{We also have (where commutators are taken in the differential crossed module $\gl(\V\btn {\cal W})$, constructed in  subsection \ref{dcmcc}).}

\begin{Lemma} {If $s$ and $t$ are degree-1 maps of $\V$ or ${\cal W}$ (according to the context) we have:}
\label{B}
\begin{align*}
 [(s \btn 1 ) , (t \btn 1)]   & = ([s ,t]) \btn 1, \\
 [(1 \btn s  ) ,( 1 \btn t )] & = 1 \btn ([s,t]) , \\
 [ t \btn 1, 1 \btn s] & = t  \btn \b(s) - \b(t) \btn s= \b'( t \btn s) \, .
\end{align*}
\end{Lemma}
And of course if $f$ and $g$ are chain maps $[f\btn 1, 1 \btn g]=0$.


\subsubsection{Tensor products of categorical representations.}

Given representations $\rho$ and $\sigma$ of the differential crossed module $(\d\colon \e \to \g,\t)$ in the chain complexes $\V$ and $\Wc$, the tensor product representation $\rho\btn\sigma$ is the representation in  $\V \btn \Wc$ such that:
$$(\rho \btn \sigma)^X_0=\rho^X_0 \btn 1 + 1 \btn \rho^X_0, $$
and also (up to 2-homotopy) 
$$(\rho \btn \sigma)^\xi_1=\rho^\xi_1 \btn 1 +1 \btn \sigma^\xi_1$$

Let us see that we have indeed defined a categorical representations. Given Lemma \ref{A}, the only complicated identity to check is:
$$[\rho^\xi_1 \btn 1 +1 \btn \sigma^\xi_1,\rho^\zeta_1 \btn 1 +1 \btn \sigma^\zeta_1]= \rho^{[\xi,\zeta]}_1 \btn 1 +1 \btn \sigma^{[\xi,\zeta]}_1 \, ,$$
up to 2-homotopy. This follows directly from Lemma \ref{B}.

We can iterate the construction of the tensor product of chain complexes, which yields the following construction: Let $\V^k=(V^k_i,\d)$, where $k=1,2,\dots,n$ be chain complexes. We define a chain complex $\V^1 \btn \V^2 \btn \ldots \btn \V^n=(K_i,\d)$, where $K_i$ is the tensor product 
$$K_i=\bigoplus_{i_1+i_2+\dots i_n= i} V^1_{i_1} \tn V^2_{i_2} \tn \ldots \tn V^n_{i_n},$$
with 
\begin{multline}
\d \big(v^1_{i_1} \tn v^2_{i_2}  \tn \dots \tn v^n_{i_n}\big) = 
\d \big(v^1_{i_1}\big) \tn v^2_{i_2}  \tn \dots \tn v^n_{i_n} + (-1)^{i_1} v^1_{i_1} \tn \d \big( v^2_{i_2} \big)  \tn \dots \tn v^n_{i_n} + \\
+ \ldots + (-1)^{(i_1+i_2+\dots+ i_{n-1})} v^1_{i_1} \tn v^2_{i_2}  \tn \dots \tn \d \big( v^n_{i_n} \big) 
\end{multline}
Given any parenthesization $\big(\V^1 \btn \V^2 \btn \ldots \btn  \V^n\big)^P$ of $\V^1 \btn \V^2 \btn \ldots \btn \V^n$, making it the iteration of $(n-1)$  two-fold  tensor products  of chain complexes, the obvious map 
$$I_P \colon \big(\V^1 \btn \V^2 \btn \ldots\btn \V^n\big)^P \to \V^1 \btn \V^2 \btn \ldots \btn \V^n$$
is an isomorphism of chain-complexes (on the nose). It is an instructive exercise to prove this for $n=3$, where the only possible  parenthesizations of $\V_1 \btn \V_2 \btn \V_3$ are  $(\V_1 \btn \V_2) \btn \V_3 $ and $\V_1 \btn (\V_2 \btn \V_3)$.

If $f^k\colon \V^k \to \V^k$ are maps of degree $m_k$ ($k=1,\dots, n$), the tensor product $f^1 \btn f^2 \btn \dots \btn f^n$ is
\begin{equation}
(f^1 \btn f^2 \btn \dots \btn f^n) (x_{i_1}^1 \otimes x_{i_2}^2 \otimes \dots \otimes x_{i_n}^n) = \chi(\{m_k\},\{i_k\}) f^1(x_{i_1}^1) \otimes f^2(x_{i_2}^2) \otimes \dots \otimes f^n(x_{i_n}^n)
\end{equation}
where 
$$ \chi(\{m_k\},\{i_k\}) =i_1(m_2+\dots +m_n) +i_2(m_3 + \dots + m_n) +\dots + i_{n-1} m_n \, . $$

Given any parenthesization $\big(	\V^1 \btn \V^2 \btn \ldots \btn   \V^n\big)^P$ of $\V^1 \btn \V^2 \btn \ldots\btn   \V^n$, we can perform the iterated tensor product $(f^1\btn \dots \btn f^n)_P$. If follows easily that  
$$I_P\circ (f^1\btn \dots \btn f^n)_P=(f^1\btn \dots \btn f^n)\circ I_P \, .$$

Easy calculations show that given categorical representations $\rho_i$ in $\V^i$, $i=(1,\ldots ,n)$, of $\mathfrak{G}=(\d:\h \to \g,\t)$ there is a tensor product categorical representation $\rho_1 \btn \rho_2 \btn \dots \btn \rho_n$ in $\V^1 \btn \V^2 \btn \dots \btn \V^n$: 
$$(\rho_1 \btn \rho_2 \btn \dots \btn \rho_n)^X_0 = (\rho_1)^X_0 \btn 1 \btn \ldots \btn 1 + 1 \btn  (\rho^X_2)_0 \btn 1 \btn \ldots \btn 1 + \dots +1 \btn \ldots \btn 1  \btn (\rho_n)^X_0$$
and also (up to 2-homotopy) 
$$(\rho_1 \btn \rho_2 \btn \dots \btn \rho_n)^\xi_1 = (\rho_1)^\xi_1 \btn 1 \btn \ldots \btn 1 + 1 \btn  (\rho_2)_1^\xi \btn 1 \btn \ldots \btn 1 + \dots + 1 \btn \ldots \btn 1 \btn (\rho_n)^\xi_1 \, . $$
Given any parenthesization of $\V_1 \btn \V_2 \btn \dots \btn \V_n$ the map 
$$I_P \colon (\V^1 \btn \V^2 \btn \dots \btn \V^n)_P \to V^1 \btn \V^2 \btn \dots \btn \V^n $$ 
is an isomorphism of categorical representations.


\subsection{Infinitesimal {2-R}-matrices}\label{defir}
Note that given a chain complex $\V$ and a positive integer $n$ there exists a representation of $S_n$ by {chain map}s $\V^{\btn n} \to \V^{\btn n}$. For a transposition $\tau_{a(a+1)}$ the associated map has the form: $$ x_1 \tn \dots \tn x_a \tn x_{a+1}\tn \dots\tn x_n \mapsto (-1)^{[x_a][x_{a+1}]} x_1 \tn \dots \tn x_{a+1} \tn x_{a} \tn \dots\tn x_n,$$ 
where $[x_a]$ and $[x_{a+1}]$ denote the degrees of $x_a$ and $x_{a+1}$. In this section we will always consider $\V^{\btn n}$ to be provided with this action; recall the construction in subsection \ref{FC}.

Let $\mathfrak{G}=(\d\colon \h \to \g,\t)$ be a differential crossed module. Let also $\V$ be a long chain complex where $\mathfrak{G}$ has a categorical representation $\rho$. Let us address $\gl(\V^{\btn n})$-valued $2$-connection $(A,B)$, of the type mentioned in Corollary \ref{2chfl}. This can be done universally in the differential crossed module $\mathfrak{G}$, as we explain now, thanks to the results presented in subsections \ref{seccatreps} and \ref{sectn} on (tensor products of) categorical representations.

The following $\g$-modules will be useful for such universal description. For a generic $k\leq n$ consider
\begin{equation}
\label{Uk}
\mathfrak{U}^{(k)}=(\e \tn \g \tn \ldots \tn\g) \oplus (\g \tn \e \tn \g \ldots \tn\g) \oplus \ldots \oplus (\g \tn \ldots \tn \g \tn \e) 
\end{equation}
where each tensor product has $k$ factors. The vector space $\mathfrak{U}^{(k)}$ corresponds to the penultimate vector space in the $k$-fold tensor product $\btn_{i=1}^k\,\, (\d\colon \h \to \g)$. We have natural maps
\begin{equation}
\label{dhat}
\hat \d\colon \mathfrak{U}^{(k)} \to \g^{\tn k} 
\end{equation}
defined, according to the decomposition \eqref{Uk}, as
$$ \hat{\d}= \d \tn \id \tn \ldots \tn\id+{\rm cyclic}$$
that is, on each summand we act with $\d$ on the $\e$ copy and we leave unchanged the other factors. 

Next, we use the categorical representation $\rho=(\rho_1,\rho_0):\mathfrak{G}\rightarrow\gl(\V)$ to associate to elements of $\mathfrak{U}^{(k)}$ (resp. $\g^{\tn k}$) homotopies in $\gl^1(\V^{\btn n})$ (resp. chain maps in $\gl^0(\V^{\btn n})$), whenever $k\leq n$. For simplicity, we denote all these maps - we call them \textit{insertion maps} - with the same symbol $\phi$. For every 
$$ \bigoplus_{i=1}^k \, u_{i_1}\tn \ldots \tn u_{i_k} \in \mathfrak{U}^{(k)}$$ 
and $\{a_1,\ldots ,a_k\}\subset \{1,\ldots ,n\}$ we define 
$$\phi_{a_1\ldots a_k} :\mathfrak{U}^{(k)} \rightarrow \gl^1(\V^{\btn n})$$ 
as
\begin{equation}
\label{phiUk}
\phi_{a_1\ldots a_k}( \bigoplus_{i=1}^k u_{i_1}\tn \ldots \tn u_{i_k})= \sum_{i=1}^k \, \id\btn\ldots \btn \rho(u_{i_1})\btn\ldots \btn\rho(u_{i_k})\btn\ldots\btn\id
\end{equation}
where in every summand we inserted the $\rho$ image of $u_{i_r}$ in the $a_r^{th}$ factor of the tensor product as the only non-trivial entries. The definition of
$$ \phi_{a_1\ldots a_k}: \g^{\tn k} \rightarrow \gl^0(\V^{\btn n}) $$
is similar.

\begin{Lemma}
\label{gmodphi}
The insertion maps 
\begin{equation*}
\phi_{a_1\ldots a_k}: \mathfrak{U}^{(k)} \rightarrow \gl^1(\V^{\btn n}) \, , \qquad \phi_{a_1\ldots a_k}: \g^{\tn k} \rightarrow \gl^0(\V^{\btn n})
\end{equation*}
are $\g$-module maps.
\end{Lemma}
\begin{Proof}
This easily follows from the definition of $\phi$ and the fact that $\rho$ intertwines the $\g$ action.
\end{Proof} \\

\begin{Lemma}
\label{dphi}
We have the following commutative diagram of $\g$-modules:
\begin{equation}
\label{dphicd}
\xymatrix{
\mathfrak{U}^{(k)}\ar[d]_-{\hat\d}\ar[rr]^-{\phi_{a_1\ldots a_k}} & & \gl^1(\V^{\btn n})\ar[d]^-{\b} \\
\g^{\tn k}\ar[rr]^-{\phi_{a_1\ldots a_k}} & & \gl^0(\V^{\btn n})
} 
\end{equation}
\end{Lemma}
\begin{Proof}
Also this property is easily verified from the definition of $\phi$ and the intertwining property $\b \rho_1 = \rho_0 \d$ of $\rho=(\rho_1,\rho_0)$.
\end{Proof} \\

Insertion maps explicitly depend on the categorical representation $\rho$, but the ones mapping into $\gl^1(\V^{\btn n})$ have a fixed contribution to their kernel due to the equivalence relation between homotopies, see Corollary \ref{simp}. For this reason we prefer to remove from the beginning elements in $\mathfrak{U}^{(k)}$ which are systematically mapped into homotopies equivalent to zero. This is achieved introducing the following subspaces: for a fixed pair of distinct indices $(r,s)\subset(1,\ldots ,k)$ and generic elements $X_{i_1}\ldots X_{i_{k-2}}\in\g$ and $w,w' \in \e$ we define
\begin{equation*}
\mathfrak{T}_{(r,s)}^{(k)} = 
\mbox{ span } \left\{ X_{i_1} \tn \ldots \tn \d w \tn \ldots \tn w' \tn \ldots \tn X_{i_{k-2}} - X_{i_1} \tn \ldots \tn w \tn \ldots \tn \d w' \tn \ldots \tn X_{i_{k-2}} \right\} \subset\mathfrak{U}^{(k)} 
\end{equation*} 
where $w$ and $w'$ are respectively in the $r^{th}$ and $s^{th}$ factor of the tensor product. We then sum over all possible pairs of distinct indices $(r,s)\in (1,\ldots ,k)$ to obtain the sub-vector space
$$\Tk  := \sum_{(r,s)} \mathfrak{T}_{(r,s)}^{(k)} \subset \mathfrak{U}^{(k)} \, .$$ 
It is clear that the $\phi$ image of $\Tk$ in $\gl^1(\V^{\btn n})$ is $2$-homotopic to the zero morphism by Corollary \ref{simp}.

\begin{Lemma}
\label{Uquot}
The quotients $\Uk:=\mathfrak{U}^{(k)}/\mathfrak{T}^{(k)} $ are $\g$-modules.
\end{Lemma}
\begin{Proof}
The $\g$-module structure induced from $\mathfrak{U}^{(k)}$ is well defined since $\mathfrak{T}^{(k)}$ is stable for $\g$-action. We prove it explicitly only for the first generator of $\mathfrak{T}_{(1,2)}^{(3)}$: for every $X,Y\in\g$ and $w,w'\in\e$ we have
\begin{equation*}
\begin{split}
Y\t (\d w\tn w'\tn X - w\tn \d w'\tn X) = & \; \d(Y\t w)\tn w' \tn X - Y\t w\tn \d w'\tn X \, + \d w\tn Y\t w' \tn X \, \\
 & \; - w \tn \d (Y\t w') + \d w\tn w' \tn [Y,X] - w\tn \d w' \tn [Y,X]\,,
\end{split}
\end{equation*}
where we used that $\d$ is a $\g$-module map.
\end{Proof} \\

We denote with $\bar{\phi}$ the insertion maps induced on the quotients $\bar{\mathfrak{U}}^{(k)}$. They are $\g$-module maps thanks to Lemma \ref{Uquot}, and they satisfy the intertwining property of diagram \eqref{dphicd} since $\hat\d(\Tk) = 0$. {Note also that $\hat{\d}\colon {\mathfrak{U}}^{(k)} \to \g^{\tn k}$ descends to $\bar{\mathfrak{U}}^{(k)}$.}

\begin{Definition}[Infinitesimal {2-R}-matrix]
\label{irmatrix}
A (non-symmetric) infinitesimal {2-R}-matrix is given by a symmetric tensor $r \in \g\tn \g$ and $P,Q \in \bar{\mathfrak{U}}^{(3)}$ such that:
\begin{equation}
 \hat{\d}(P)=[r_{12}+r_{13},r_{23}] \quad \quad \an \quad \quad \hat{\d}(Q)=[r_{12}+r_{13},r_{23}]
\end{equation}
(in $\g \tn \g \tn \g$) and also:
\begin{equation}
\label{condflun}
\begin{split}
& r_{14}\act (Q_{213} + P_{234}) + (r_{12} + r_{23} + r_{24}) \act Q_{314} - (r_{13}+r_{34})\act Q_{214} = 0 \\
& r_{24}\act (P_{123} + P_{134}) + (r_{12} + r_{14} + r_{13}) \act Q_{324} - (r_{23}+r_{34})\act P_{124} = 0 \\
& r_{23}\act (Q_{214} + Q_{314}) + r_{14}\act (Q_{324} + P_{234} - P_{123}) = 0 \\
& r_{13}\act (P_{124} + Q_{324}) + r_{24}\act (Q_{314} + P_{134} - Q_{213}) = 0 \\
& r_{34}\act (Q_{213} + Q_{214}) + (r_{12} + r_{23} + r_{24}) \act P_{134} - (r_{13}+r_{14})\act P_{234} = 0 \\
& r_{34}\act (P_{123} + P_{124}) + (r_{12} + r_{13} + r_{14}) \act P_{234} - (r_{24}+r_{23})\act P_{134} = 0 \\
\end{split}
\end{equation}
These last relations are to hold in $\bar{\mathfrak{U}}^{(4)}$.
\end{Definition}
Note that if $r=\sum_i s_i \otimes t_i$, then $r_{12}$, $r_{13}$ and $r_{23}$ are elements of $\U(\g)\tn \U(\g) \tn \U(\g)$ defined as in \eqref{rab}. Therefore for example $[r_{12},r_{23}]=\sum_{i,j} s_i \tn [t_i, s_j] \tn t_j \in \g \tn \g \tn \g.$ On the other hand,  if $P=\sum_{j} A_j \tn B_j \tn C_j$, then:
$$r_{14} \t P_{234}=\sum_{i,j} s_i \tn   A_j \tn B_j \tn t_i \t C_j \in  \bar{\mathfrak{U}}^{(4)}, $$
where $\t$ can either be, depending on $j$, the adjoint action of $\g$ on $\g$ or the given action of $\g$ on $\h$.

\begin{Definition}[Totally symmetric infinitesimal {2-R}-matrix]
A totally symmetric infinitesimal {2-R}-matrix is an infinitesimal {2-R}-matrix where 
\begin{align}
\label{Aa} 
P&=Q &P_{123}+P_{231}+P_{312}&=0, &P_{123}&=P_{132} \, .
\end{align}
Therefore, by the calculations in subsection \ref{FC}, a totally symmetric infinitesimal {2-R}-matrix is given by a symmetric tensor $r\in \g \tn \g$, an element $P\in\bar{\mathfrak{U}}^{(3)} $, with $\hat{\d}(P)=[r_{12}+r_{13},r_{23}]$, such that the two last equations of \eqref{Aa} are satisfied, and moreover (these relations are to hold in $\bar{\mathfrak{U}}^{(4)}$):
\begin{equation}
\label{sym2R}
\begin{split}
& r_{14}\act (P_{213} + P_{234}) + (r_{12} + r_{23} + r_{24}) \act P_{314} - (r_{13}+r_{34})\act P_{214} = 0 \\
& r_{23}\act (P_{214} + P_{314}) -r_{14}\act (P_{423} +P_{123}) = 0 \, .
\end{split}
\end{equation}
In other words, adding conditions \eqref{Aa} one can show that \eqref{condflun} reduce to \eqref{sym2R}.
\end{Definition}
\begin{Example}
Choose a Lie algebra $\g$ and a tensor $r\in \g \tn \g$. Consider the crossed module given by the identity map $\g \stackrel{\id}{\to} \g$ and the adjoint action of $\g$ on $\g$. Then $\bar{\mathfrak{U}}^{(n)}=\g^{\tn n}$. Therefore by the discussion above $(r,[r_{12}+r_{13},r_{23}])$ is a totally symmetric  infinitesimal {2-R}-matrix.
\end{Example}

From the discussion in subsections \ref{FC} and \ref{dcm} we have the following result.
\begin{Theorem}
\label{ABflat}
Let $(r,P,Q)$ be an infinitesimal {2-R}-matrix on the differential crossed module $\mathfrak{G}=(\d:\h \to \g,\t)$. Consider a categorical representation of $\mathfrak{G}$ on a long complex of vector spaces $\V$. Consider the $\gl(\V^{\btn n})$-valued $2$-connection $(A,B)$ on the configuration space $\C(n)$ defined as
\begin{align}
A &= \sum_{a<b} \w_{ab} \, \bar{\phi}_{ab}(r) \, .\\
B &= \sum_{a<b<c} \w_{ab}\wedge\w_{ac} \, \bar{\phi}_{bac}(Q) + \w_{ab}\wedge \w_{bc} \, \bar{\phi}_{abc}(P)
\end{align}
Then $(A,B)$ is a flat 2-connection. Moreover if $(r,P,Q)$ is totally symmetric the $2$-connection is invariant under the action of the symmetric group $S_n$ and its {two-dimensional} holonomy descends to a {two-dimensional} holonomy in $\C(n)/S_n$ with values in the associated  Lie crossed module $\GL(\V^{\btn n})$.
\end{Theorem}

\section*{Acknowledgements}
The first author (LSC) was partially supported by INdAM, through `Borse di studio estero 2010-2011'. The second author (JFM) was  partially supported by CMA/FCT/UNL, under the project
PEst-OE/MAT/UI0297/2011. This work was partially supported by FCT (Portugal) through the projects
PTDC/MAT/098770/2008 
and PTDC/MAT/101503/2008. 

\end{document}